\documentclass[12pt,eadjoint tfnpsf]{article}

\catcode`\@=11
\@addtoreset{equation}{section}

\global\arraycolsep=1pt

\usepackage{amsbsy,amssymb,latexsym,amsfonts, amsmath}
\usepackage{mathrsfs}
\usepackage{graphicx}

\RequirePackage[dvips,usenames]{color}
\definecolor{fireblick}{rgb}{0.698039,0.133333,0.133333}

\newcommand{\beq}{\begin{equation}}
\newcommand{\eeq}{\end{equation}}
\newcommand{\bea}{\begin{eqnarray}}
\newcommand{\eea}{\end{eqnarray}}

\newcommand{\CB}{{\mathcal B}}

\newcommand{\CE}{{\mathcal E}}
\newcommand{\CF}{{\mathcal F}}
\newcommand{\CH}{{\mathcal H}}

\newcommand{\CN}{{\mathcal N}}
\newcommand{\CO}{{\mathcal O}}
\newcommand{\CP}{{\mathcal P}}

\newcommand{\CW}{{\mathcal W}}

\renewcommand\Im{{\mathrm{Im}}}

\def\Tr{\mathop{\rm Tr}}

\newcommand{\fundl}
{\setlength{\unitlength}{0.6pt}
\begin{picture}(10,10)
\put(0,0){\line(1,0){10}}
\put(0,10){\line(1,0){10}}
\put(0,0){\line(0,1){10}}
\put(10,0){\line(0,1){10}}
\end{picture}}

\newcommand{\anti}
{\setlength{\unitlength}{0.55pt}
{\begin{picture}(16,22)
\put(4,-4){\line(0,1){20}}
\put(4,16){\line(1,0){10}}
\put(4,6){\line(1,0){10}}
\put(4,-4){\line(1,0){10}}
\put(14,-4){\line(0,1){20}}
\end{picture}}}

\newcommand{\symm}
{\setlength{\unitlength}{0.55pt}
{\begin{picture}(24,14)
\put(0,0){\line(0,1){10}}
\put(0,10){\line(1,0){20}}
\put(0,0){\line(1,0){20}}
\put(10,0){\line(0,1){10}}
\put(20,0){\line(0,1){10}}
\end{picture}}}

\renewcommand{\thefootnote}{\fnsymbol{footnote}}

\begin{document}
%
%
\begin{titlepage}

\begin{flushright}
\normalsize
~~~~
YITP-10-45\\
June, 2010 \\
\end{flushright}

\vspace{80pt}

\begin{center}
{\LARGE Deformed Prepotential, Quantum Integrable System}\\
\vspace{10pt}
{\LARGE and Liouville Field Theory}\\
\end{center}

\vspace{25pt}

\begin{center}
{
Kazunobu Maruyoshi\footnote{e-mail: maruyosh@yukawa.kyoto-u.ac.jp}
and Masato Taki\footnote{e-mail: taki@yukawa.kyoto-u.ac.jp}
}\\
%
\vspace{15pt}
%
\it Yukawa Institute for Theoretical Physics, Kyoto University, Kyoto 606-8502, Japan\\
\end{center}
%
\vspace{20pt}
\begin{center}
Abstract\\
\end{center}
We study the dual descriptions recently discovered for the Seiberg-Witten theory 
in the presence of surface operators.
The Nekrasov partition function for a four-dimensional $\CN=2$ gauge theory with a surface operator
is believed equal to the wave-function of the corresponding integrable system, or the Hitchin system,
and is identified with the conformal block with a degenerate field via the AGT relation.
We verify the conjecture by showing 
that the null state condition leads to the Schr\"odinger equations of the integrable systems.
Furthermore, we show that the deformed prepotential emerging from the period integrals of the principal function
corresponds to monodromy operation of the conformal block.
We also give the instanton partition functions for the asymptotically free $SU(2)$ gauge theories
in the presence of the surface operator via the AGT relation.
We find that these partition functions involve the counting of two- and four-dimensional instantons.

\vfill

\setcounter{footnote}{0}
\renewcommand{\thefootnote}{\arabic{footnote}}

\end{titlepage}

\section{Introduction}
\label{sec:intro}
  The $\CN=2$ supersymmetric gauge theories provide us an interesting framework
  where symmetry constrains non-perturbative dynamics 
  and is powerful enough to lead to exact result of the low energy effective action \cite{SW, SW2}.
  It has been known that this exact solution possesses an interpretation 
  in terms of integrable systems \cite{Gorsky, MW, NT, DW, IM1, IM2, Gorsky2}.
  The Seiberg-Witten curve of an $\CN=2$ supersymmetric gauge theory is identified 
  with the spectral curve of the integrable system.
  The low energy prepotential of gauge theory can therefore be obtained by the period integrals of
  the meromorphic one form on the spectral curve.
  
  This interpretation was further sophisticated recently by \cite{NS} to the relation between the prepotential
  with a nonzero deformation parameter, which Nekrasov's partition function \cite{Nekrasov} gives, 
  and quantization of the integrable system.
  The nonzero deformation parameter plays the role of the Planck constant for the quantum integrable system.
  See also \cite{NW, Orlando:2010uu, Kozlowski:2010tv}.
  It was also proposed that the above deformed prepotential can be obtained 
  by the similar procedure to Seiberg-Witten theory, i.e. period integrals, 
  where the meromorphic one form is changed to the quantum corrected one \cite{MM1}.
  The exact WKB solution for the Schr\"odinger equation of the integrable model gives this quantum one form. 
  This was further studied in \cite{MM2, Popolitov, He:2010xa}.
  
  Meanwhile, a new insight has been added to $\CN=2$ gauge theories.
  In \cite{Gaiotto}, it was found that
  the compactification of the six-dimensional (2,0) $A_{N-1}$ theory on a Riemann surface 
  leads to a colossal class of $\CN=2$ superconformal $SU(N)$ quiver gauge theories.
  The Seiberg-Witten curve for a theory in this class is realized as a $N$-tuple cover of this Riemann surface.
  Then, a remarkable relation 
  between the Nekrasov partition function \cite{Nekrasov} of $\CN=2$ superconformal $SU(2)$ gauge theory
  and the conformal block of two-dimensional Liouville field theory was proposed
  by Alday, Gaiotto and Tachikawa \cite{AGT}.
  (We refer to this as AGT relation.)
  This conjecture was generalized to the relation between the asymptotically free $SU(2)$ gauge theories
  and irregular conformal blocks \cite{Gaiotto:2009ma, Marshakov:2009gn},
  and also to the higher rank case \cite{Wyllard, MM}.
  
  The AGT relation is very useful to analyze various observables in gauge theories.
  In particular, the partition function in the presence of a surface operator is identified 
  with the conformal block with an additional insertion of the degenerate field
  in the Liouville theory \cite{AGGTV}.
  It was also conjectured that the Wilson and t' Hooft loop operators correspond
  to the monodromy operations for the degenerate field inserted conformal block 
  along some cycles of the Riemann surface \cite{AGGTV, DGOT}.
  (See also \cite{DMO, Kozcaz:2010af, GMN2}.)
  
  In this paper, we relate the quantization of the integrable system 
  with the insertions of the surface and the Wilson-t' Hooft loop operators 
  in the gauge theory partition function, concentrating on the Liouville and $SU(2)$ gauge theories.
  It is well-known that a conformal block with degenerate fields satisfies a differential equation \cite{BPZ}.
  For the degenerate field $\Phi_{2,1}(z)$ with momentum $-\frac{b}{2}$ (or $\Phi_{1,2}(z)$ with $- \frac{1}{2b}$)
  which is the case we will consider in this paper, the differential equation is quadratic
  since the null state condition is also quadratic in the Virasoro generators $((L_{-1})^2 + b^2 L_{-2})\Phi_{2,1}(z)=0$.
  By taking the limit where one deformation parameter goes to zero,
  we interpret the reduced differential equation as the Schr\"odinger equation of the associated integrable system.
  The Hamiltonian of the system can therefore be read off from the Liouville theory consideration.
  We confirm that the Hamiltonian obtained from the torus conformal block
  which corresponds to the $\CN=2^*$ gauge theory is that of the elliptic Calogero-Moser system.
  We also consider degenerate field insertion in the conformal block 
  corresponding to the $SU(2)$ gauge theory with four flavors,
  and in the irregular conformal blocks corresponding to asymptotically free theories.
  These cases also support the conjecture that
  a Schr\"odinger system is associated with a gauge theory.
  
  Based on these observations, we find that 
  the proposal in \cite{MM1} that the deformed prepotential would be obtained 
  from the solution of the Schr\"odinger equation 
  is equivalent to expected monodromies of the conformal block with the degenerate field inserted. 
  The $A$- and $B$-cycle monodromies are expected to be
  the phase shift by the expectation value of vector multiplet scalar $a$ 
  and the shift of the vev $a$ as $a \rightarrow a + \epsilon_2$ respectively.
  We will see that assuming the proposal \cite{MM1} leads to the monodromy conditions stated above.
  Conversely, the monodromy conditions verify the proposal.
  While we see this correspondence at lower orders in the Planck constant (deformation parameter),
  we expect that this relation is valid even at higher orders.
  
  We study the details of the irregular Virasoro conformal blocks with the degenerate field
  which are expected  to be equal to the Nekrasov partition functions
  for $SU(2)$ asymptotically free theories in the presence of a surface operator.
  By expanding it in the Verma module,
  we obtain the  Nekrasov-like partition function which has two expansion parameters $\Lambda$ and $z$.
  We can recast it into the expansion in terms of the
  four-dimensional instanton factor $\Lambda^4$ and two-dimensional one $\Lambda^2z$.
  The irregular Virasoro conformal block with the degenerate field
  therefore describes the two- and four-dimensional instanton counting for the surface operator.
  This result supports our expectation that the insertion of the degenerate field
  leads to the Nekrasov instanton partition function in the presence of a surface operator
  for superconformal theories, and also for asymptotically free theories.
  
  The organization of this paper is as follows.
  In section \ref{sec:quant}, we consider the proposal in \cite{MM1} 
  which relates the quantization of the integrable system with the deformation of the prepotential of the gauge theory.
  We will analyze $\CN=2^*$ gauge theory and $SU(2)$ gauge theory with four flavors as examples.
  In section \ref{sec:Liouville}, we show that the null state condition for the (irregular) conformal blocks
  in the presence of the degenerate field implies the Schr\"odinger equations for the associated integrable systems.
  We also see the equivalence between the proposal stated in section \ref{sec:quant} 
  and monodromy operation of the conformal block with the degenerate field.
  In section 4, we study the structure of the irregular Virasoro conformal blocks with the degenerate field
  in the perspective of the instanton counting.
  We conclude with discussions in section \ref{sec:conclusion}.
  In appendix \ref{sec:Nekrasov}, we briefly review the Nekrasov partition function.
  In appendix \ref{sec:ellipticintegral}, the explicit calculation of the energy eigenvalue in section \ref{sec:quant}
  will be presented.
  In appendix \ref{sec:cb}, we consider the action of the degenerate field on the Verma module.
  
  While preparing this paper, \cite{Teschner, Alday:2010vg, Dimofte:2010tz} 
  which have some overlap with this paper appeared.

\section{Quantum Integrable Systems}
\label{sec:quant}
  It is already known that the Seiberg-Witten solutions for an $\mathcal{N}=2$ gauge theory
  is described by a classical integrable system.
  The family of the Seiberg-Witten curves is realized as 
  the family of the energy-levels of the Hamiltonian for the system.
  Then quantum integrable systems can be associated with some extension of the Seiberg-Witten theory,
  such as Nekrasov's theory of instanton counting.
  It was conjectured in \cite{NS} that ``$\epsilon$ deformation" of the prepotential, 
  which defined by
    \bea
    \CF(\epsilon_1)
     =     \lim_{\epsilon_2 \rightarrow 0} (- \epsilon_1 \epsilon_2) Z_{{\rm Nek}},
           \label{deformedprepotential0}
    \eea
  where $Z_{{\rm Nek}}$ is the Nekrasov partition function \cite{Nekrasov},
  are related to the quantization of the integrable system
  (see appendix \ref{sec:Nekrasov} for a review of the Nekrasov partition function). 
  In \cite{MM1} it was proposed that the deformed prepotential can also be obtained 
  by considering the Schr\"odinger equation of the system.
  
  In this section we will see that this proposal works 
  by evaluating the deformed prepotential from several integrable models.
  First of all, let us briefly see the proposal.
  We consider the Schr\"odinger equation of a model 
    \bea
    \CH \Psi^{(0)}(z)
     =     E \Psi^{(0)}(z),
           \label{Schrodinger}
    \eea
  where Hamiltonian is $\CH = - \epsilon_1^2 \partial^2_z +  V(z; \epsilon_1)$.
  The meaning of the subscript of the wave-function will be clear in next section.
  (In a few examples below, the right hand side would be further multiplied by a $z$-dependent factor.
  However, this will not affect the generic analysis below.)
  We then write the wave-function in terms of a one form $P(z) dz$
    \bea
    \Psi^{(0)}(z)
     =     \exp \left( - \frac{1}{\epsilon_1} \int^z P(z'; \epsilon_1) dz' \right).
           \label{P}
    \eea
  This is just the exact WKB ansatz, since the one form is expanded in a power series of the Planck constant $\epsilon_1$.
  The claim \cite{MM1} is that the integrals of the one form over the $A$- and $B$-cycles can be written as
    \bea
    2 \pi i \hat{a}_i(E; \epsilon_1)
    &=&    \oint_{A^i} P(z; \epsilon_1) dz,
           \nonumber \\
    \frac{1}{2} \frac{\partial \hat{\CF}}{\partial \hat{a}_i} (E; \epsilon_1)
    &=&    \oint_{B^i} P(z; \epsilon_1) dz,
           \label{AB}
    \eea
  and that after eliminating $E$ by using the first equation, $\hat{\CF}$ coincides with
  the deformed prepotential of an $\CN=2$ gauge theory $\CF(\epsilon_1)$ (\ref{deformedprepotential0}).
  In other words, this means that the monodromies of the wave-function around the $A$- and $B$-cycles are
    \bea
    \Psi^{(0)}(z + A^i)
    &=&    \exp \left( - \frac{2 \pi i \hat{a}_i}{\epsilon_1} \right) \Psi^{(0)}(z), 
           \nonumber \\
    \Psi^{(0)}(z + B^i)
    &=&    \exp \left(- \frac{1}{2 \epsilon_1} \frac{\partial \hat{\CF}}{\partial \hat{a}_i} \right) \Psi^{(0)}(z),
           \label{MM}
    \eea
  where $i=1, \ldots, g$ ($g$ is genus of the curve).
  
  Note that in order for the claim to be meaningful, we have to specify which potential corresponds 
  to particular $\CN=2$ gauge theory.
  Before mentioning it, we first give a generic prescription to obtain the deformed prepotential.
  By substituting (\ref{P}) into the Schr\"odinger equation, we obtain
    \bea
    - P^2 + \epsilon_1 P' + V(z; \epsilon_1) 
     =     E,
           \label{equationP}
    \eea  
  where $P$ and $V$ are expanded in $\epsilon_1$ as
    \bea
    P(\epsilon_1; z)
     =     \sum_{k=0}^\infty \epsilon_1^k P_k(z), ~~
    V(z;\epsilon_1)
     =     \sum_{k=0}^\infty \epsilon_1^k V_k(z).
    \eea
  At lower orders, (\ref{equationP}) gives the following recursion relations:
    \bea
    - P_0^2 + V_0
    &=&    E,
           \nonumber \\
    - 2 P_0 P_1 + P'_0 + V_1
    &=&    0,
           \nonumber \\
    - 2 P_0 P_2 - P_1^2 + P'_1 + V_2
    &=&    0.
    \eea
  Therefore, we obtain the expansion of the one form
    \bea
    P_0
     =     \sqrt{V_0 - E}, ~~~
    P_1
     =     \frac{1}{2 P_0} (P'_0 + V_1),
           ~~~
    P_2
     =     \frac{1}{2 P_0} (P'_1 - P_1^2 + V_2 ),
           \label{P0P1P2}
    \eea
  and so on.
  In the first equation, we have chosen the plus sign.
  As we will see that in explicit examples, a contour integral of the $\epsilon_1$-deformed one form, $\oint Pdz$,
  is written as an action of an operator $\hat{\CO}$ on the classical (zero-th order) one: 
  $\oint P = \hat{\CO} \oint P_0 = (1 + \epsilon_1 \hat{\CO}_1 + \epsilon_1^2 \hat{\CO}_2 + \ldots) \oint P_0$.
  
  The remaining task is the following 
  (we focus on the case of $SU(2)$ gauge group where the curve $P_0^2 = V_0 - E$ is genus one, in what follows):
   \begin{enumerate}
   \item calculate the $A$-cycle integral: $2 \pi i a(E) = \oint_A P_0 dz$,
   \item obtain $\frac{\partial \CF}{\partial a} (E)$ by calculating the $B$-cycle integral, 
         or by using a known result of $\CF(a)$
         (from, e.g. the Nekrasov partition function) and 
         then substituting $a(E)$ obtained in the step $1$ into it:
         $\frac{\partial \CF}{\partial a} (a) = \frac{\partial \CF}{\partial a} (a(E))$ as in \cite{MM2}, 
   \item act the operator $\hat{\CO}$ on both ones obtained in the steps 1 and 2:
           \bea
           \hat{a}(E; \epsilon_1)
            =     \hat{\CO} \Big[ a(E) \Big], ~~~
           \frac{\partial \hat{\CF}}{\partial \hat{a}} (E; \epsilon_1)
            =     \hat{\CO} \left[ \frac{\partial \CF}{\partial a} (E) \right],
           \eea
   \item rewrite $E$ in terms of $\hat{a}$: $E = E(\hat{a})$ and
         substitute it into $\frac{\partial \hat{\CF}}{\partial \hat{a}} (E; \epsilon_1)$.
         By integrating over $\hat{a}$, we obtain $\hat{\CF}(\hat{a}; \epsilon_1)$.
   \end{enumerate}
  Finally, we compare this result with the deformed prepotential (\ref{deformedprepotential0}).
  
  In the gauge theory point of view, integrable systems described above correspond 
  to gauge theories with $SU(2)$ gauge group.
  At classical level (zero-th order in $\epsilon_1$), the correspondence with Seiberg-Witten theory
  can be accomplished by the identification of the curve $P_0^2 = V_0 - E$ with the Seiberg-Witten curve.
  More precisely, the potential is identified with $\phi_2^{{\rm SW}}(z)$ 
  in the Seiberg-Witten curve $x^2 = \phi_2^{{\rm SW}}$
  obtained from the M-theory construction \cite{Witten, Gaiotto, GMN}
  without the term depending on the Coulomb moduli $u$.
  The Coulomb moduli $u$ plays the role of the energy eigenvalue of the Schr\"odinger equation (\ref{Schrodinger}).
  The quantum correction promotes the Seiberg-Witten theory 
  to the deformation of the prepotential in the presence of $\Omega$-background.
  
  It is known that for pure Yang-Mills and $\CN=2^*$ theories, the corresponding integrable systems
  are the periodic Toda and the elliptic Calogero-Moser theories respectively \cite{Gorsky, MW, IM1}.
  (For $SU(2)$ case, the periodic Toda is precisely the sine-Gordon model.)
  Gauge theories with fundamental hypermultiplets have also been considered in \cite{Gorsky2}.
  
  The above proposal has been verified in \cite{MM1} by calculating lower order $\epsilon_1$ expansion 
  of $\hat{\CF}(\hat{a})$ in the sine-Gordon model, that is the model with the potential
    \bea
    V(\theta)
     =     \Lambda^2 \cos \theta
    \eea
  and by comparing it with the deformed prepotential.
  Note that the coordinate $\theta$ was introduced by $z = e^{i \theta}$.
  
  There is another interesting property in this method.
  It is known that the derivative of the prepotential with respect to the gauge coupling constant
  corresponds to the Coulomb moduli $u$ \cite{Matone, STY, EY}.
  This is even true in the $\epsilon$-deformed case as found in \cite{Flume, Fucito}.
  It would therefore be natural to expect that the energy obtained in the last step $E = E(\hat{a})$ coincides 
  with the derivative of the deformed prepotential $\CF(a; \epsilon_1)|_{a=\hat{a}}$ 
  with respect to the gauge coupling constant.
  Namely, the energy $E(\hat{a})$ can be identified with the ``deformed" Coulomb moduli.
  We can check this property in the sine-Gordon model above.
  
  To see further validity of the claim and also the property stated just above, 
  let us consider a few examples below.

\subsubsection*{$\CN=2^*$ gauge theory}
  Let us consider the integrable system corresponding to the $\CN=2^*$, $SU(2)$ gauge theory,
  that is the $SU(2)$ gauge theory with an adjoint hypermultiplet with mass $m$.
  The Hamiltonian which we will consider below is that of elliptic Calogero-Moser system
  $\CH = - \epsilon_1^2 \partial^2_z + m (m - \epsilon_1) \CP(z)$ where $\CP(z)$ is the Weierstrass elliptic function 
  (with periods $\pi$ and $\pi \tau$).
  In other words, the potential is given by
    \bea
    V(\epsilon_1; z)
     =     m (m - \epsilon_1) \CP(z).
    \eea 
  We will confirm this choice from the Liouville theory point of view in section \ref{sec:Liouville}.
  In this case, $V_0 = m^2 \CP(z)$, $V_1 = - m \CP(z)$ and all the others vanish.
  Therefore, $P(z)$ can be written as
    \bea
    P_0
     =     \sqrt{m^2 \CP(z) - E}, ~~~
    P_1
     =     \frac{1}{2 P_0} (P'_0 - m \CP(z)), ~~~
           \ldots.
    \eea
  
  In the following, we will consider the contour integral of $P$.
  By explicit computation, the contour integral of $P_1$ 
  is simplified as
    \bea
    \oint P_1 dz
    &=&  - \frac{m}{2} \oint \frac{\CP(z)}{P_0} dz
     =   - \frac{1}{2} \frac{\partial}{\partial m} \oint P_0 dz.
    \eea
  Therefore, the operator $\hat{\CO}$ becomes in this case,
    \bea
    \hat{\CO}
     =     1 - \frac{\epsilon_1}{2} \frac{\partial}{\partial m} + \CO(\epsilon_1^2).
           \label{operatorN=2*}
    \eea
  
  We first note that at leading order in $\epsilon_1$, the curve $P_0^2 = m^2 \CP(z) - E$ is the Seiberg-Witten curve
  of the $\CN=2^*$, $SU(2)$ gauge theory \cite{DW} (see also \cite{Gaiotto}).
  Therefore, the $A$- and $B$-cycle integrals lead to the gauge theory prepotential \cite{MNW, Fucito, FL}.
  We then consider the periods of the quantum corrected one form $P dz$.
  For the first order calculation in $\epsilon_1$, 
  a powerful simplification occurs because the first order term in $\hat{\CO}$ is merely the derivative 
  with respect to the mass parameter.
  Indeed, the actions of the operator $\hat{\CO}$ at the first order are simply the following shift
    \bea
    \hat{a}(E)
     =     a (E)|_{m^{k} \rightarrow m^{k} - k \epsilon_1 m^{k-1}/2}, 
           ~~~
    \frac{\partial \hat{\CF}}{\partial \hat{a}}(E)
     =     \frac{\partial \CF}{\partial a}(E)|_{m^{k} \rightarrow m^{k} - k \epsilon_1 m^{k-1}/2}.
           \label{actionON=2*}
    \eea
  Then, we solve the first equation for $E$: $E = E(\hat{a}; \epsilon_1) = E(a)|_{a\rightarrow \hat{a},~
  m^{k} \rightarrow m^{k} - k \epsilon_1 m^{k-1}/2}$, 
  where $E(a)$ is the classical expression of the energy.
  By substituting this into the second equation of (\ref{actionON=2*}), we obtain
  $\frac{\partial \hat{\CF}}{\partial \hat{a}}(\hat{a}, \epsilon_1) 
  = \frac{\partial \CF}{\partial a}(a)|_{a\rightarrow \hat{a},~
  m^{k} \rightarrow m^{k} - k \epsilon_1 m^{k-1}/2} + \CO(\epsilon_1^2)$.
  Therefore, at this order, the deformed prepotential is simply given by 
  $\hat{\CF}(\hat{a}, \epsilon_1) 
  = \CF(a)|_{a\rightarrow \hat{a}, ~
  m^{k} \rightarrow m^{k} - k \epsilon_1 m^{k-1}/2} + \CO(\epsilon_1^2)$.
  Actually, we can calculate the instanton part
    \bea
    \hat{\CF}_{{\rm inst}}(\hat{a})
    &=&     \frac{m^4}{2\hat{a}^2} q + \frac{m^4 (96\hat{a}^4 - 48\hat{a}^2 m^2 + 5 m^4)}{64\hat{a}^6} q^2
          + \ldots \nonumber \\
    & &   - \left[ \frac{m^3}{\hat{a}^2}q 
          + \frac{m^3(48 \hat{a}^4 - 36 \hat{a}^2 m^2 + 5 m^4) }{16 \hat{a}^6} q^2 + \ldots \right] \epsilon_1
          + \CO(\epsilon_1^2),
            \label{deformedprepotentialN=2*}
    \eea
  where $q = e^{2 \pi i \tau}$.
  This agrees with the deformed prepotential which obtained from the Nekrasov partition function of the $\CN=2^*$ theory
  with $\epsilon_2 = 0$ while keeping $\epsilon_1$ finite.
  This observation is quite simple, but already non-trivial result.
  
  The property stated before was that the energy can be expressed as the derivative of the prepotential.
  The above argument also shows this at least at the first order in $\epsilon_1$,
  once we verify the property at the classical level.
  For completeness let us check this.
  In order to get the expression of $E$, we have to compute the $A$-cycle integral of the one form $P_0 dz$.
  As analyzed in \cite{FL}, it is convenient to introduce $M = m^2/a^2$ and write $2 \pi i a = \oint_A P_0 dz$ as
    \bea
    \pi 
     =     \oint_A \sqrt{\CE - \frac{M}{4}\CP(z)},
           \label{EN=2*}
    \eea
  where $\CE = E/4a^2$.
  By solving this, we obtain the series $\CE = \sum_{i=0} \CE_i(q) M^i$ 
  where $\CE_0 = 1$ and the higher coefficients are functions only of the coupling $q$.
  We will give explicit expressions of lower order $\CE_i(q)$ in appendix \ref{sec:ellipticintegral}.
  It follows from these that 
    \bea
    E
     =     4 \left( a^2 - \frac{m^2}{12} + \frac{m^2 (4 a^2 + m^2)}{2 a^2}q 
         + \frac{m^2 (192 a^6 + 96 m^2 a^4 - 48 m^4 a^2 + 5 m^6)}{32 a^6}q^2 + \ldots \right).
    \eea
  This can be written as
    \bea
    E
     =     4 q \frac{\partial}{\partial q} \left( \CF(a) - 2 m^2 \ln \eta(\tau) \right)
     \equiv
           4 q \frac{\partial \tilde{\CF}(a)}{\partial q},
    \eea
  where $\eta(\tau)$ is Dedekind eta function.
  Note that the one-loop contribution does not appear since this is $q$ derivative.
  It deserves mentioning that the difference between $E$ and $\frac{\partial \CF}{\partial \ln q}$
  has already observed in \cite{SW2, Dorey, Fucito}.
  This is due to the difference $\tilde{u} = \langle \Tr \phi^2 \rangle + \ldots$ 
  where $\tilde{u}$ corresponds here to $E$, the variable in the curve,
  and $\langle \Tr \phi^2 \rangle$ to the derivative of the prepotential.
  
  Since we know that the action of $\hat{\CO}$ on $\CF(a)$ gives rise to $\hat{\CF}(\hat{a})$ 
  as in (\ref{deformedprepotentialN=2*}), 
  the corrected energy is evaluated as
    \bea
    E(\hat{a})
     =     4 q \frac{\partial}{\partial q} \left( \hat{\CF}(\hat{a}) - 2(m^2 - m \epsilon_1) \ln \eta(\tau) \right),
    \eea
  at the first order in $\epsilon_1$.

\subsubsection*{$SU(2)$ gauge theory with $N_f = 4$}
  Then, let us consider a more complicated theory.
  The potential corresponding to the $SU(2)$ gauge theory with four flavors is 
    \bea
    V
    &=&    \frac{\tilde{m}_1^2 - \frac{\epsilon_1^2}{4}}{z^2} + \frac{m_0 (m_0 - \epsilon_1)}{(z - 1)^2}
         + \frac{m_1 (m_1 - \epsilon_1)}{(z - q)^2}
         - \frac{m_0 (m_0 - \epsilon_1) + m_1 (m_1 - \epsilon_1) + \tilde{m}_1^2 - \tilde{m}_0^2}{z(z-1)}.
           \nonumber \\
           \label{VNf4}
    \eea
  We will also see the origin of this potential in section \ref{sec:Liouville}.
  It is easy to see the leading order potential
    \bea
    V_0
     =     \frac{\tilde{m}_1^2}{z^2} + \frac{m_0^2}{(z - 1)^2} + \frac{m_1^2}{(z - q)^2}
         - \frac{m_0^2 + m_1^2 + \tilde{m}_1^2 - \tilde{m}_0^2}{z(z-1)}
    \eea
  is almost the Seiberg-Witten curve \cite{Gaiotto} except for the Coulomb moduli term
  $- \frac{(1-q)u}{z(z-1)(z-q)}$.
  Also, the higher order terms are
  $V_1 = - \frac{1}{2} \left( \frac{\partial}{\partial m_0} + \frac{\partial}{\partial m_1} \right) V_0$
  and $V_2 = - \frac{1}{4z^2}$. 
  Similar to the above case, lower order terms in $P$ can be evaluated as
    \bea
    P_0
     =     \sqrt{V_0 - \frac{(1-q)E}{z(z-1)(z-q)}}, 
           ~~~~
    P_1
     =   - \frac{1}{2} \left( \frac{\partial}{\partial m_0} + \frac{\partial}{\partial m_1} \right) P_0
         +  d \left( \frac{1}{2} \log P_0 \right), 
           \ldots
           \label{P0P1}
    \eea
  Note that $P_0 dz$ is the same as the Seiberg-Witten one form \cite{Gaiotto, EM, EM2}.
  Therefore, at the classical level, the gauge theory prepotential can be obtained 
  from its $A$- and $B$-cycle integrals.
  
  The first order correction $P_1$ is mass derivative of $P_0$.
  Therefore, the simplification similar to the $\CN=2^*$ theory occurs.
  We can check that the deformed prepotential can be obtained by the action of the operator $\hat{\CO}$.
  For instance, the prepotential is computed from the Nekrasov instanton partition function 
  (see appendix \ref{sec:Nekrasov} for a review)
    \bea
    \CF_{{\rm inst}}(a)
     =     \frac{1}{2a^2} \left( a^4 + (m_0^2 + m_1^2 - \tilde{m}_1^2 - \tilde{m}_0^2) a^2
         + (m_0^2 -  \tilde{m}_0^2)(m_1^2 - \tilde{m}_1^2) \right) q + \CO(q^2).
    \eea
  By acting $\hat{\CO}$, we obtain the first order $\epsilon_1$ deformation:
    \bea
    - \frac{\epsilon_1}{2a^2} 
    \left( (m_0 + m_1) a^2 - m_0 \tilde{m}_1^2 + m_0^2 m_1 + m_0 m_1^2 - \tilde{m}_0^2 m_1 \right),
    \eea
  which agrees with the deformed prepotential as in appendix \ref{sec:Nekrasov}.
  We have verified this for lower instanton expansion.
  
  Let us here check that $E(\hat{a})$ is given by the derivative of the deformed prepotential 
  with respect to the gauge coupling constant.
  For simplicity, let us consider the equal mass case where $m_0 = m_1 = m$ and $\tilde{m}_0 = \tilde{m}_1=0$.
  (Actually, these parameters are the linear combinations of the masses of four hypermultiplets 
  and the equal mass case corresponds to the above choice (\ref{massSU(2)}).)
  At the classical level, we can compute the $A$-cycle integral of $P_0 dz$ and obtain
  (see appendix \ref{sec:ellipticintegral}) for a detailed calculation)
    \bea
    E
    &=&    a^2 - m^2 + \frac{a^4 + 2 m^2 a^2 + m^4}{2a^2}q
         + \frac{13a^8 + 36m^2 a^6 + 22m^4 a^4 - 12m^6 a^2 + 5m^8}{32a^6} q^2 + \CO(q^3).
           \nonumber \\
    \eea
  We can easily check that this is the derivative of the prepotential $\CF(\epsilon_1 = 0)$ with respect to $\ln q$
  (with additional terms $a^2 - m^2$ whose origin will be found in subsection \ref{subsec:degenerate}).
  
  Then, we apply the operator $\hat{\CO}$.
  As we have seen in the last paragraph, this action is, at the first order in $\epsilon_1$, merely 
  the shift of the mass parameter.
  Thus, at this order, it is easy to obtain 
  $E(\hat{a}) = \frac{\partial \hat{\CF} (\hat{a})}{\partial \ln q} - m^2 + m \epsilon_1$.
  We expect that this relation is satisfied even at higher orders.
  This relation will become important in the subsequent section.
  
  In summary, we have seen that the deformed prepotential with finite $\epsilon_1$
  is obtained from the quantization of the Schr\"odinger system at the first order in $\epsilon_1$
  in several examples above.
  We expect that this is still satisfied for higher order terms.
  The reason of this will be explained in the subsequent section, by relating this problem 
  to monodromy operation of the correlation function with the degenerate field insertion in the Liouville theory.

\section{Liouville Field Theory and Deformed Prepotential}
\label{sec:Liouville}
  So far, we have seen that the deformed prepotentials can be obtained from the quantum integrable systems.
  In that, the choice of the potential was somewhat heuristic.
  In this section, we will see that the form of the potential can be dictated 
  from the degenerate field insertion in the conformal block.
  After a brief review of the AGT relation, we consider the differential equations 
  which are satisfied by the conformal blocks (the correlation function in the case on a torus) 
  with one degenerate field in subsection \ref{subsec:degenerate}.
  Irregular conformal blocks \cite{Gaiotto:2009ma} which have been identified 
  with the Nekrasov partition functions of the asymptotically free $SU(2)$ gauge theories 
  will be analyzed in subsection \ref{subsec:irregular}.
  Then, we relate the proposal in the previous section 
  to the monodromy condition on the degenerate field inserted conformal block in subsection \ref{subsec:monodromy}.

\subsection{Degenerate field and surface operator}
\label{subsec:degenerate}
  A class of four-dimensional $\CN=2$ superconformal gauge theory can be obtained 
  from six-dimensional $(2,0)$ theory on a Riemann surface of genus $g$ with $n$ punctures \cite{Witten, Gaiotto}.
  The AGT relation \cite{AGT} relates the Nekrasov instanton partition function 
  of $\CN=2$ superconformal $SU(2)$ gauge theory
  with the conformal block of the Liouville field theory on the Riemann surface:
    \bea
    Z_{{\rm inst}}(a_p, m_i; \epsilon_1, \epsilon_2)
     =     \CB(\alpha_p^{int}, \alpha_i; b)
           \label{AGT}
    \eea
  where the primary fields with Liouville momenta $\alpha_i$ are inserted at the points of the punctures.
  We note that when discussing the conformal block, 
  we have to specify the choice of pants decomposition of the Riemann surface.
  This corresponds to the weak coupling description of the gauge theory.
  Various possible pants decompositions are related by the mapping class group of the Riemann surface
  which is interpreted as S-duality transformation in the gauge theory point of view \cite{Gaiotto}.
  
  The instanton partition functions of the gauge theories analyzed in the previous section,
  the $SU(2)$ gauge theory with four flavors and the $\CN=2^*$ $SU(2)$ gauge theory, 
  are identified with the conformal blocks on a sphere with four punctures 
  and on a torus with one puncture.
  
  More precise identification of the parameters in the relation (\ref{AGT}) are as follows. 
  The vacuum expectation value of the vector multiplet scalar $a$ is related 
  with the primary field in intermediate line by
    \bea
    \Delta^{int}_p
     =     \frac{Q^2}{4} - \frac{a_p^2}{\hbar^2},
           \label{Coulomb}
    \eea
  where $p =1, \ldots, 3g - 3 + n$ and $Q = b + 1/b$ ($b$ is the Liouville parameter).
  The number of the $SU(2)$ gauge groups is equal to $3g - 3 + n$.
  The mass parameter, roughly speaking, corresponds to the field inserted at the puncture.
  Note however that this parameter is the one associated with an $SU(2)$ flavor symmetry as (\ref{massSU(2)})
  and also some of $m$'s are shifted by $Q/2$ from the gauge theory values \cite{AGT}, 
  e.g., for the four point conformal block on a sphere,
    \bea
    \Delta_{\alpha_1}
     =     \frac{Q^2}{4} - \frac{\tilde{m}_0^2}{\hbar^2}, ~~~
    \Delta_{\alpha_2}
     =     \frac{m_0}{\hbar} (Q - \frac{m_0}{\hbar}), ~~
    \Delta_{\alpha_3}
     =     \frac{m_1}{\hbar} (Q - \frac{m_1}{\hbar}), ~~
    \Delta_{\alpha_4}
     =     \frac{Q^2}{4} - \frac{\tilde{m}_1^2}{\hbar^2}.
           \label{mass}
    \eea
  The deformation parameters are related with the parameter of the Liouville theory $b$ via
    \bea
    \epsilon_1
     =     \frac{\hbar}{b}, ~~
    \epsilon_2
     =     \hbar b.
           \label{epsilon}
    \eea
  Finally, the coupling constants defined at UV region $q_i = e^{2 \pi i \tau_i}$ 
  of the $SU(2)^{3g - 3 + n}$ gauge groups are identified with the complex structures of the Riemann surface.
  
  We have to note the definition of the conformal block.
  The chiral half of the full Liouville correlation function can be written as, 
  e.g., for the four-point function, 
    \bea
    \left< V_{\alpha_1}(\infty) V_{\alpha_2} (1) V_{\alpha_3}(q) V_{\alpha_4} (0) \right>
     =     q^{\Delta^{int} - \Delta_3 - \Delta_4} \CB(\alpha^{int}, \alpha_i; b),
           \label{conformalblock}
    \eea
  up to the DOZZ factors.
  As in (\ref{AGT}), $\CB$ was identified with the Nekrasov instanton partition function
  and is expanded in $q$ as $\CB = 1 + \CO(q)$.
  However, in the rest of this subsection, we call the left hand side of (\ref{conformalblock}) as conformal block.
  
  The prepotential of the gauge theory is obtained by taking the limit of the Nekrasov partition function
  \cite{Nekrasov, Nekrasov:2003rj}.
  Therefore, by making use of the AGT relation, it can be extracted from the conformal block:
  $\CF = \lim_{\epsilon_1, \epsilon_2 \rightarrow 0} (- \epsilon_1 \epsilon_2) \CB$.
  The deformed prepotential (\ref{deformedprepotential0}) can also be obtained, via the AGT relation, as follows:
    \bea
    \CF(\epsilon_1)
     =     \lim_{\epsilon_2 \rightarrow 0} (- \epsilon_1 \epsilon_2) \CB.
           \label{deformedprepotential}
    \eea
  Also, as pointed out in \cite{AGT}, 
  the ``quantum" Seiberg-Witten curve is given by the insertion of the energy-momentum tensor:
    \bea
    x^2
     =     \phi_2 (z)
     \equiv
           \frac{\langle T(z) \prod_{i=1}^n V_{\alpha_i} (z_i) \rangle}{\langle \prod_{i=1}^n V_{\alpha_i} (z_i) \rangle},
           \label{quantizedSW}
    \eea
  where $T(z) = \sum_{n\in \mathbb{Z}} L_n/z^{n+2}$.
  Indeed, this reduces in the limit $\epsilon_{1,2} \rightarrow 0$ to the Seiberg-Witten curve in \cite{Gaiotto}
    \bea
    x^2
     =     \lim_{\epsilon_{1,2} \rightarrow 0} \frac{1}{\epsilon_1 \epsilon_2} \phi_2 (z)
     =     \phi_2^{{\rm SW}}(z).
    \eea
  
  We consider an additional insertion of the degenerate field in the conformal block.
  We concentrate on the degenerate field $\Phi_{2,1}$ (or $\Phi_{1,2}$), 
  which is the operator with Liouville momentum $- \frac{b}{2}$ (or $-\frac{1}{2b}$):
    \bea
    \Psi(a_p, z)
     =     \left< \Phi_{2,1}(z) \prod_{i=1}^n V_{\alpha_i} (z_i) \right>.
    \eea
  Due to the null field condition $(b^2 L_{-2} + (L_{-1})^2) \Phi_{2,1} (z) = 0$, 
  $\Psi$ satisfies the second order differential equation \cite{BPZ}.
  In \cite{AGGTV}, $\Psi$ was identified with the surface operator insertion 
  in the Nekrasov instanton partition function.
  Compared with the conformal block, $\Psi$ depends on $z$.
  
  Let us consider the limit where $\epsilon_2 \rightarrow 0$ while $\epsilon_1$ fixed.
  In this limit the dependence of $z$ would appear in the subleading term in $\epsilon_2$:
    \bea
    \Psi
     =     \exp \left( - \frac{1}{\epsilon_1 \epsilon_2} (\CF(\epsilon_1)
         + \epsilon_2 \CW(z; \epsilon_1) + \CO(\epsilon_2^2)) \right),
           \label{Psiepsilon2}
    \eea
  where the first term is the deformed prepotential (\ref{deformedprepotential})
  with additional terms due to the difference similar to (\ref{conformalblock}).
  By solving the differential equation as in section \ref{sec:quant}, it is possible to obtain $\CW(z; \epsilon_1)$.
  
  We can also consider the insertion of the degenerate field $\Phi_{1,2}$.
  However, this leads to the similar equation with $\epsilon_1$ and $\epsilon_2$ exchanged.
  Therefore, we concentrate on $\Phi_{2,1}$ insertion in what follows.
  We will derive the differential equations in the following examples.
  
\subsubsection*{Sphere with four punctures}
  To begin with, let us consider the conformal block on a sphere with four punctures.
  Before considering the degenerate field insertion, we see that the insertion of the energy-momentum tensor
  (\ref{quantizedSW}), for the four-point conformal block, gives rise to
    \bea
    \phi_2(z) 
     =     \sum_{i=1}^4 \left( \frac{\Delta_i}{(z - z_i)^2}
         + \frac{1}{z - z_i} \frac{\partial}{\partial z_i} \right) \langle \prod_{i=1}^4 V_{\alpha_i}(z_i) \rangle /
           \langle \prod_{i=1}^4 V_{\alpha_i}(z_i) \rangle.
           \label{SWNf4}
    \eea
  Since the four-point conformal block satisfies
  $0 = \sum_{i=1}^4 \hat{\Lambda}_a \left< \prod_{i=1}^4 V_{\alpha_i}(z_i) \right>$ ($a= -1, 0, 1$),
  where 
    \bea
    \hat{\Lambda}_{-1}
     =     \sum_{i=1}^4 \frac{\partial}{\partial z_i},~~~
    \hat{\Lambda}_{0}
     =     \sum_{i=1}^4 (z_i \frac{\partial}{\partial z_i} + \Delta_i), ~~~
    \hat{\Lambda}_{1}
     =     \sum_{i=1}^4 (z_i^2 \frac{\partial}{\partial z_i} + 2 \Delta_i z_i), 
           \label{lambda}
    \eea
  we can rewrite $z_i$ derivative in (\ref{SWNf4}) in terms of a derivative with respect to only one position, say $z_3$.
  Then, we choose the position of the puncture as $z_1 = \infty$, $z_2 = 1$, $z_3 = q$ and $z_4 = 0$.
  After some algebra, we obtain 
  $\phi_2(z) 
  = \hat{\phi}_2(z) \left< \prod_{i=1}^4 V_{\alpha_i}(z_i) \right>/\left< \prod_{i=1}^4 V_{\alpha_i}(z_i) \right>$ with
    \bea
    \hat{\phi}_2
    &=&    \frac{\Delta_4}{z^2} + \frac{\Delta_3}{(z - q)^2} + \frac{\Delta_2}{(z-1)^2}
           \nonumber \\
    & &  - \frac{1}{z(z-1)(z -q)} \left( (1-q) \frac{\partial}{\partial \ln q} 
         + (z - q) (\sum_{i=2}^4 \Delta_i - \Delta_1) \right).
           \label{phi2Nf4}
    \eea
  This reproduces the Seiberg-Witten curve of the $SU(2)$ gauge theory with four flavors 
  in the limit where $\epsilon_{1,2} \rightarrow 0$.
  
  Then, we consider the insertion of the degenerate field $\Phi_{2,1}(z)$ in the conformal block:
  $\Psi(z) = \left< \Phi_{2,1}(z) \prod_{i=1}^4 V_{\alpha_i} (z_i) \right>$.
  To study the constraint equation which the null state condition $((L_{-1})^2 + b^2 L_{-2}) \Phi_{2,1} = 0$ implies,
  we compute the action of these Virasoro operators on the degenerate field.
  Let us introduce the action of the Virasoro operators as
    \begin{align}
    T(w)\Phi(z)
     =    \sum_{n\in \mathbb{Z}}\frac{1}{(w-z)^{n+2}} L_n\Phi(z).
    \end{align}
  By differentiating the above OPE with respect to $z$, 
  we obtain the OPE between descendant field $\partial \Phi$ and the energy-momentum tensor
    \begin{align}
    T(w)\partial \Phi(z)
     =    \frac{2\Delta}{(w-z)^{3}}\Phi(z) + \frac{n+1}{(w-z)^2}\partial \Phi(z)
        + \frac{1}{w-z}\partial^2 \Phi(z) + \cdots .
    \end{align}
  The action of $(L_{-1})^2$ on the primary is thus given by the differential operation
  ${(L_{-1})}^2\, \Phi(z) = \partial^2_z \Phi(z)$.
  By combining this and the action of $L_{-2}$, 
  the null state condition leads to the following differential equation
    \bea
    0
     =     \left[ b^{-2} \partial^2_z + \sum_{i=1}^4 \left( \frac{\Delta_i}{(z - z_i)^2}
         + \frac{1}{z - z_i} \frac{\partial}{\partial z_i} \right) \right] \Psi(z).
    \eea
  Since $\Psi$ satisfies the similar relation to (\ref{lambda}) 
  (in this case the indices run from $1$ to $4$ and also $z$), 
  we obtain
    \bea
    \label{BPZ}
    0
     =     \left[ b^{-2} \partial^2_z + \hat{\phi}_2
         - \frac{1}{z(z-1)(z-q)} \left( (2z - 1) \frac{\partial}{\partial z} + \Delta \right) \right] \Psi(z),
    \eea
  where $\hat{\phi}_2$ is the same as (\ref{phi2Nf4}) 
  and $\Delta$ is the conformal dimension of the degenerate field: $\Delta = - \frac{1}{2} - \frac{3b^2}{4}$.
  
  We translate the parameters to the gauge theory ones by using (\ref{Coulomb}) -- (\ref{epsilon}).
  We then take a limit where $\epsilon_2 \rightarrow 0$, where the conformal dimensions behave as 
  $\Delta_{\alpha_1} \hbar^2 = \frac{\epsilon_1^2}{4} - \tilde{m}_0^2 + \CO(\epsilon_2)$,
  $\Delta_{\alpha_2} \hbar^2 = m_0 (\epsilon_1 - m_0) + \CO(\epsilon_2)$ and so on.
  Thus, we obtain
    \bea
    0
     =     \left( - \epsilon_1^2 \partial^2_z + V(z; \epsilon_1) 
         - \frac{(1-q)}{z(z-1)(z-q)} \frac{\partial \CF(\epsilon_1)}{\partial \ln q}\right) \Psi^{(0)}(z),
           \label{differentialS04}
    \eea
  where $V(z; \epsilon_1)$ is the same one as (\ref{VNf4})
  and $\Psi^{(0)} = \lim_{\epsilon_2 \rightarrow 0} \Psi / \langle \prod V_{\alpha_i}(z_i) \rangle$.
  We have used the asymptotics (\ref{Psiepsilon2}) and therefore $\frac{\partial \CF(\epsilon_1)}{\partial \ln q} 
  = a^2 - m_1^2 - \tilde{m}_1^2 + m_1 \epsilon_1 + \frac{\partial \CF_{{\rm inst}}(\epsilon_1)}{\partial \ln q}$,
  where the first four terms come from $-(\Delta^{int} - \Delta_3 - \Delta_4)$
  because of the definition of the conformal block (\ref{conformalblock}).
  Note that in the case $\epsilon_1 = 0$, the second and the third terms in the right hand side of (\ref{differentialS04}) 
  is $\phi_2^{{\rm SW}}$ except that the Coulomb moduli is changed to $\frac{\partial \CF}{\partial \ln q}$.
  Therefore, we have ``derived" the Schr\"odinger equation analyzed in the previous section.
  The wave-function and the degenerate conformal block $\Psi^{(0)}$ play the similar roles.
  
  Note that while in the analysis in section \ref{sec:quant}, the energy was {\it a priori} unknown parameter,
  we ``know" its value here by using the AGT relation.
  This will be an important point in subsection \ref{subsec:monodromy}.
  
  We could extend this argument to a generic quiver gauge theory.
  The Schr\"odinger equation obtained by inserting the degenerate field then involves one variable $z$.
  While a quiver gauge theory corresponds to an integrable system with many canonical variables,
  it is expected that the Hamiltonian of the Schr\"odinger equation is not for this many body system itself, but
  for the wave-function associated with the Baker-Akhiezer function related with the system \cite{DV, Schiappa:2009cc}.
  This leads to the Schr\"odinger equation with single variable $z$.
  
  It may be helpful to comment on the relation to the Hitchin system.
  In \cite{Feigin:1994in} it was shown 
  that the Hitchin system associated with a four-punctured sphere is the Gaudin model of $SL(2)$-type.
  Let $\Phi^a_i$ be the representations of the generators of $SL(2)$:
     \bea
    \Phi^{-}_i
     =  \frac{\partial}{\partial t_i},~~~
    \Phi^{0}_i
     =    t_i \frac{\partial}{\partial t_i} + j_i, ~~~
    \Phi^{+}_i
     = t_i^2 \frac{\partial}{\partial t_i} + 2 j_i t_i.
    \eea
 We introduce the Higgs field
 \bea
 \Phi=\left(\begin{array}{cc}\Phi^{0} & \Phi^{+} \\
 \Phi^{-} & -\Phi^{0}\end{array}\right),
 \eea
 where
 \bea
 \Phi^a=\sum_{i=1}^4 \frac{\Phi^{a}_i }{z-z_i}.
 \eea
 Then the quantum spectral curve
  \bea
 \det \left( \Phi(z)-\epsilon \partial_z\right)=0
 \eea
 implies the so-called Gaudin Hamiltonians $H_i$:
  \bea
 \epsilon^2\frac{\partial^2}{\partial z^2}-\sum_i \left( \frac{j_i(j_i+1)}{(z-z_i)^2} +\frac{H_i}{z-z_i}\right)=0
 \eea 
 This equation resembles the differential equation (\ref{BPZ}).
 Let us give a rough sketch of the connection between them.
 In the limit $b\to 0$ the last term of (\ref{BPZ}) reduces to the accessory parameter terms ${C_i}/{(z-z_i)}$.
 Since in \cite{Ribault:2005wp} the accessory parameters $C_i$ are identified with the eigenvalues of the Gaudin Hamiltonians $H_i$
 through the $H_3^+$-Liouville correspondence,
 the Schr\"odinger system for $SU(2)$ gauge theory with four flavors corresponds to the quantization
 of the Hitchin system for the four-punctured sphere.
 See \cite{Teschner} for recent development.

\subsubsection*{Torus one-point function}
  The torus one-point conformal block was identified with 
  the Nekrasov instanton partition function of the $\CN=2^*$, $SU(2)$ gauge theory \cite{AGT}:
  $Z_{{\rm inst}}^{\CN=2^*}(a, m; \epsilon_1, \epsilon_2) = \CB(\alpha^{int}, \alpha; b)$, where 
  $\left< V_{\alpha}(0) \right>_\tau^{{\rm full}} = \int d \alpha^{int} \ldots |q^{\Delta^{int} - \frac{c}{24}} \CB|^2$.
  In order to obtain the differential equation, 
  we consider the full correlation function with one additional degenerate field insertion:
    \bea
    \left< \Phi_{2,1}(z) V_{\alpha'}(0) \right>_\tau^{{\rm full}}.
    \label{degeneratetorus}
    \eea
  Note that we have shifted the momentum of the external field in (\ref{degeneratetorus}) as $\alpha' = \alpha + b/2$
  due to the degenerate field insertion.
  
  Due to the null field condition, 
  the correlation function (\ref{degeneratetorus}) satisfies the following differential equation \cite{EO}: 
    \bea
    & &    \left( - b^{-2} \partial_{z}^2 - \eta_1 + (\zeta(z) - 2 \eta_1 z) \partial_{z}
         - \Delta_{\alpha'} (\CP(z) + 2\eta_1) \right)
           \left< \Phi_{2,1}(z) V_{\alpha'}(0) \right>_\tau^{{\rm full}}
           \nonumber \\
    & &    ~~~~~~~~~~
     =     \left( \frac{2 i}{\pi} \frac{\partial}{\partial \tau}
         + \eta_1 - \frac{1}{2 \pi \Im \tau} \right) \left< \Phi_{2,1}(z) V_{\alpha'}(0) \right>_\tau^{{\rm full}} ,
    \eea
  where $\zeta(z)$ and $\eta_1$ are defined in appendix \ref{sec:ellipticintegral}.
  We write the correlation function as \cite{FL}
    \bea
    \left< \Phi_{2,1}(z) V_{\alpha'}(0) \right>_\tau^{{\rm full}}
     =     \left( \vartheta_1(z|\tau) \right)^{b^2/2} \left( \eta(\tau) \right)^{2 \Delta_{\alpha'} - 1 - 2 b^2}
           \Psi(z|\tau).
           \label{Psitorus}
    \eea
  Here $\vartheta_1(z|\tau)$ is elliptic theta function.
  In terms of $\Psi$, the differential equation gets simplified as
    \bea
    \left( - b^{-2} \partial^2_z - (\Delta_{\alpha'} - \frac{1}{2} - \frac{b^2}{4}) \CP(z)
    - \eta_1 (1 + \frac{3b^2}{2}) + \frac{1}{2 \pi \Im \tau} \right) \Psi(z|\tau)
     =    \frac{2 i}{\pi} \frac{\partial}{\partial \tau} \Psi(z|\tau).
          \label{fulldifferentialN=2*}
    \eea
  
  By translating the parameters to the gauge theory ones, as in the case of a sphere with four punctures, 
  the above equation leads to
    \bea
    \left( - \epsilon_1^2 \partial^2_z +  m (m - \epsilon_1) \CP(z) + \CO(\epsilon_2) \right) \Psi(z|\tau)
     =     \frac{2 i}{\pi} \epsilon_1 \epsilon_2 \frac{\partial}{\partial \tau} \Psi(z|\tau),
    \eea
  where we have used that 
  $\Delta_{\alpha'}\hbar^2 = (m + \frac{\epsilon}{2}) (\epsilon_+ - m - \frac{\epsilon}{2})
   = m (\epsilon_1 - m) + \CO(\epsilon_2)$
  and the fact that all the terms in the left hand side in (\ref{fulldifferentialN=2*}) 
  except for $\Delta_{\alpha'}$ term and the derivative term are of order $\CO(\epsilon_2)$.
  Then, we consider the limit where $\epsilon_2 \rightarrow 0$.
  Since $\Psi$ is the correlation function, the chiral half of should behave as
    \bea
    \exp \left[ - \frac{1}{\epsilon_1 \epsilon_2} \left( a^2 \ln q + \tilde{\CF}_{{\rm inst}} (\epsilon_1)
    + \CO(\epsilon_2) \right) \right],
    \eea
  where the first term comes from $q^{\Delta^{int} - \frac{c}{24}}$ in the full correlation function.
  We claim that the leading term $\tilde{\CF}_{{\rm inst}}(\epsilon_1)$ is the same 
  as the one obtained by the same limit in the one-point conformal block:
    \bea
    \CB
     =     (\eta(\tau))^{\frac{2m(\epsilon_1 - m)}{\epsilon_1 \epsilon_2}}
           \exp \left( - \frac{1}{\epsilon_1 \epsilon_2} \left( \tilde{\CF}_{{\rm inst}}(\epsilon_1)
         + \CO(\epsilon_2) \right) \right).
           \label{torusconformalblockshift}
    \eea
  Note that we have multiplied $(\eta(\tau))^{\frac{2m(\epsilon_1 - m)}{\epsilon_1 \epsilon_2}}$ 
  in order to be consistent with (\ref{Psitorus}),
  and therefore 
  $\tilde{\CF}_{{\rm inst}}(\epsilon_1) (= \CF_{{\rm inst}}(\epsilon_1) - 2(m^2- m \epsilon_1) \ln \eta(\tau))$ 
  is different from the instanton prepotential of $SU(2)$ gauge theory.
  Note also that the first factor $\vartheta_1^{b^2/2}$ in (\ref{Psitorus}) is of order 
  $\frac{\epsilon_2^2}{\epsilon_1 \epsilon_2}$ which is irrelevant in our analysis.
  Then, by ignoring $\epsilon_2$ terms, we obtain the following differential equation:
    \bea
    \left( - \epsilon_1^2 \partial^2_z +  m (m - \epsilon_1) \CP(z) \right) \Psi^{(0)}(z|\tau)
     =     4 \frac{\partial \tilde{\CF}(\epsilon_1)}{\partial \ln q} \Psi^{(0)}(z|\tau),
           \label{differentialtorus}
    \eea
  where we have defined 
  $\frac{\partial \tilde{\CF}}{\partial \ln q} = a^2 + \frac{\partial \tilde{\CF}_{{\rm inst}}}{\partial \ln q}$, 
  including the classical part.
  The left hand side is the Hamiltonian of the elliptic Calogero-Moser system introduced in section \ref{sec:quant}.
  Note that the elliptic Calogero-Moser system is also the Hitchin system for a torus with a puncture \cite{DW}.

\subsection{Insertion of degenerate field  into irregular conformal blocks}
\label{subsec:irregular}

The original AGT relation is the map between the Virasoro conformal blocks and the instanton
partition functions for the $\mathcal{N}=2$ superconformal $SU(2)$ quiver gauge theories.
It is to be anticipated that 
we can formulate analogous relation for the $SU(2)$ asymptotically free theories \cite{Gaiotto:2009ma}.
We then have to define the ``irregular" conformal blocks in the CFT side which correspond to the ``wild"
singularities of the quadratic differentials $\phi_2^{{\rm SW}}(z)$ of these gauge theories.
Coherent states \cite{Gaiotto:2009ma} which live in the Verma module,
which are called the Gaiotto states,
are the basic building blocks of these irregular conformal blocks 
for $SU(2)$ gauge theories with $N_f = 0, 1,2,3$ flavors.

The Nekrasov instanton partition function for the $SU(2)$ pure super Yang-Mills theory is
\begin{align}
Z_{\textrm{inst}}(a ,\Lambda,\epsilon_1,\epsilon_2)
 =\sum_{k=0}^{\infty}\Lambda^{4k} Z_{k}(a; \epsilon_1,\epsilon_2)
 =\sum_{k=0}^{\infty}\frac{\Lambda^{4k}}{(\epsilon_1\epsilon_2)^{2k}} Z_{k}\left(\alpha; b \right),
\end{align}
where $\alpha=a/\sqrt{\epsilon_1\epsilon_2}$ and $b=\sqrt{\epsilon_2/\epsilon_1}$
are dimensionless parameters.
(For a moment we also use the dimensionless dynamical scale $\Lambda \rightarrow \Lambda \sqrt{\epsilon_1 \epsilon_2}$.)
See appendix \ref{sec:Nekrasov} for details of the construction of the partition function.
Notice that the $k$-instanton factor $Z_k$ is a homogeneous function with degree $-4k$.
Gaiotto found out in \cite{Gaiotto:2009ma} that the partition function is equal to a certain irregular
conformal block of the Virasoro algebra as follows
\begin{align}
Z_{\textrm{inst}}(\alpha,\,\Lambda;\,b)=\langle \Delta, \Lambda^2 |\Delta, \Lambda^2 \rangle,
\end{align}
where the conformal dimension is $\Delta(\alpha)=Q^2/4 - \alpha^2$. 
Here the two Gaiotto states $ |\Delta, \Lambda^2 \rangle$ 
are associated with the two wild singularities of the punctured sphere
on which the conformal block is defined.
The Gaiotto state $ |\Delta, \Lambda^2 \rangle= |\Delta \rangle +\cdots$
satisfies the coherent state condition
\bea
L_0  |\Delta,\Lambda^2 \rangle
 = \left(\Delta+\frac{\Lambda}{2} \frac{\partial}{\partial \Lambda}\right)|\Delta,\Lambda^2 \rangle, ~~~
L_1  |\Delta,\Lambda^2 \rangle = \Lambda^2|\Delta,\Lambda^2 \rangle,
\label{gaiottoL0}
\eea
and $L_n  |\Delta,\Lambda^2 \rangle =0$ for  $n \geq 2$,
where $|\Delta\rangle$ is the highest weight state with conformal dimension $\Delta$.

In \cite{Marshakov:2009gn, Alba:2009fp}
the explicit solution for the Gaiotto state $ |\Delta, \Lambda^2 \rangle$ 
is given in terms of the Shapovalov matrix $Q_{\Delta}(Y;Y^{\prime})$.
The Shapovalov matrix is the following Gram matrix:
\begin{align}
Q_{\Delta}(Y;Y^{\prime})=\langle \Delta |L_{Y}L_{-Y^{\prime}} |\Delta \rangle.
\end{align}
Here $Y=\{Y_1,\, Y_2,\, \cdots\}=\left[1^{m_1}2^{m_2}\cdots \right]$ is a Young diagram with $|Y|=\sum Y_i=\sum j\,m_j$ boxes,
and $L_{-Y}$ denotes $L_{-Y_l}\cdots L_{-Y_2}\cdot L_{-Y_1}$.
In \cite{Marshakov:2009gn, Alba:2009fp}
the authors proved that the following state solves the constraint equations (\ref{gaiottoL0})
\begin{align}
\label{gaiottostate}
 |\Delta,n \rangle=\sum_{|Y|=n} Q_{\Delta}^{-1}([1^n]; Y)\,L_{-Y}\,|\Delta \rangle .
\end{align}
This result also means that the existence of the Gaiotto state has been proved. 
In this way, we can rewrite the AGT relation for the pure $SU(2)$ Yang-Mills into the following form
\begin{align}
\label{GMMM}
Z_{\textrm{inst}}^{\,\,{\textrm{pure}}}(\alpha,\,\Lambda;\,b)=\sum_{n} \Lambda^{4n}\, Q_{\Delta}^{-1}([1^n]; [1^n]).
\end{align}

In order to rewrite the Nekrasov partitions for gauge theories with fundamental matters as irregular conformal blocks,
we introduce the coherent state $|\Delta,\Lambda, m \rangle $ which satisfies
\bea
L_1  |\Delta,\Lambda, m \rangle= -2m\Lambda|\Delta,\Lambda, m \rangle,~~~
L_2  |\Delta,\Lambda, m \rangle= -\Lambda^2|\Delta,\Lambda, m \rangle,
\eea
and $L_n  |\Delta,\Lambda, m \rangle =0$ for $n \geq 3$,
where $m$ 
corresponds the mass of a hypermultiplet in the gauge theory side.
By using this coherent state, 
we can recast the Nekrasov partition functions for $N_f=1,2$ theories 
in the following irregular conformal blocks \cite{Gaiotto:2009ma}
\begin{align}
&Z_{\textrm{inst}}^{N_f=1}(\alpha,\,m,\,\Lambda;\,b)
=\langle \Delta, \Lambda, m |\Delta,{\Lambda^2}/{2} \rangle,\\
&Z_{\textrm{inst}}^{N_f=2}(\alpha,\,m_1,\,m_2,\,\Lambda;\,b)
=\langle \Delta, \Lambda, m_2 |\Delta,\Lambda, m_1 \rangle.
\end{align}
See \cite{Marshakov:2009gn} for the relation to the Shapovalov matrix elements.
These non-conformal AGT relations have been proved recently in \cite{HJS},
by using the Zamolodchikov recursion relation \cite{Zamolodchikov:1985ie, Poghossian, HJS0}.

In the rest of this subsection, we study the differential equations 
which the null state condition impose on the irregular conformal blocks, as in the previous subsection.
We will follow the discussion of \cite{Awata:2009ur} 
where the case of the pure super Yang-Mills theory was discussed.

\subsubsection*{$SU(2)$ pure super Yang-Mills theory}
We study the insertion of the degenerate field $\Phi_{2,1}$ into the irregular conformal blocks.
The irregular conformal block for the pure super Yang-Mills theory 
in the presence of the degenerate field is given by
\begin{align}
\Psi(z)=\langle \Delta^\prime , \Lambda^2 |\,\Phi_{2,1}(z) | \Delta,\Lambda^2 \rangle.
\end{align}
We set $\Delta^{\prime}=\Delta(\alpha + b/4) $ and $\Delta=\Delta(\alpha -b/4)$ in accordance with the fusion rule.
In order to derive the differential equation for $\Psi$, 
we consider the insertion of the energy momentum tensor in the conformal block.
Since the higher-order Virasoro generators annihilate the Gaiotto state $ L_{n\geq 2}  |\Delta,\Lambda^2 \rangle =0$, 
we can rewrite it as follows:
\begin{align}
\label{TPhi21}
&\langle \Delta^\prime , \Lambda^2 |\, T(w)\Phi_{2,1}(z) | \Delta,\Lambda^2 \rangle\nonumber\\
& \rule{0pt}{4ex} =\sum_{n=0}^{\infty}\frac{1}{w^{n+2}}
\langle \Delta^\prime , \Lambda^2 |[L_n ,\, \Phi_{2,1}(z) ]| \Delta,\Lambda^2 \rangle\nonumber\\
&\quad  \rule{0pt}{4ex}+\frac{1}{w}\langle \Delta^\prime , \Lambda^2 |\,L_{-1}\, \Phi_{2,1}(z) | \Delta,\Lambda^2 \rangle
+\frac{1}{w^2}\langle \Delta^\prime , \Lambda^2 |\, \Phi_{2,1}(z)L_0 | \Delta,\Lambda^2 \rangle
+\frac{1}{w^3}\langle \Delta^\prime , \Lambda^2 |\, \Phi_{2,1}(z)L_1 | \Delta,\Lambda^2 \rangle\nonumber\\
& \rule{0pt}{5ex} =\left( \frac{z}{w(z-w)}\frac{\partial}{\partial z} +\frac{\Delta_{2,1}}{(w-z)^2} 
+\left(\frac{\Lambda^2}{w} +\frac{\Lambda^2}{w^3} \right)\right)\Psi(z)
+\frac{1}{w^2} 
\langle \Delta^\prime , \Lambda^2 |\,\Phi_{2,1}(z)L_0
 | \Delta,\Lambda^2 \rangle,
\end{align}
where we have used the coherent state condition (\ref{gaiottoL0}).
The following relation holds for the last term of the above equation
\begin{align}
\label{21L0}
\langle \Delta^\prime , \Lambda^2 |\,\Phi_{2,1}(z)L_0 | \Delta,\Lambda^2 \rangle
=\frac{1}{2}\left(
\frac{\Lambda}{2} \frac{\partial}{\partial \Lambda} +\Delta +\Delta^\prime -\Delta_{2,1} -z\frac{\partial}{\partial z}
\right)\Psi(z).
\end{align}
We can show this relation
by using the commutation relation $[L_0,\Phi_{2,1}(z)]=(z\partial_z +\Delta_{2,1})\Phi_{2,1}(z)$ 
and (\ref{gaiottoL0}):
\begin{align}
\frac{\Lambda}{2} \frac{\partial}{\partial \Lambda} \Psi(z)
&= \rule{0pt}{4ex}
\langle \Delta^\prime , \Lambda^2 |(L_0-\Delta^\prime)\, \Phi_{2,1}(z) | \Delta,\Lambda^2 \rangle
+\langle \Delta^\prime , \Lambda^2 |\, \Phi_{2,1}(z)(L_0-\Delta) | \Delta,\Lambda^2 \rangle
\nonumber\\
& \rule{0pt}{4ex}=-(\Delta +\Delta^\prime)\Psi(z)
+\langle \Delta^\prime , \Lambda^2 |[L_0, \Phi_{2,1}(z)] | \Delta,\Lambda^2 \rangle
+2\langle \Delta^\prime , \Lambda^2 |\, \Phi_{2,1}(z)L_0 | \Delta,\Lambda^2 \rangle
\nonumber\\
& \rule{0pt}{4ex}=
-\left(
\Delta +\Delta^\prime -\Delta_{2,1} -z\frac{\partial}{\partial z}
\right)\Psi(z)
+2\langle \Delta^\prime , \Lambda^2 |\, \Phi_{2,1}(z)L_0 | \Delta,\Lambda^2 \rangle.
 \end{align}
By substituting (\ref{21L0}) into (\ref{TPhi21}), we obtain the following expression
\begin{align}
\label{TPhi21_2}
\langle \Delta^\prime , \Lambda^2 |\, T(w)\Phi_{2,1}(z) | \Delta,\Lambda^2 \rangle
&=
\Big[ \frac{z}{w(z-w)}\frac{\partial}{\partial z} +\frac{\Delta_{2,1}}{(w-z)^2} 
+\left(\frac{\Lambda^2}{w} +\frac{\Lambda^2}{w^3} \right)
\nonumber\\
&\qquad+\frac{1}{2w^2}\left(
\frac{\Lambda}{2} \frac{\partial}{\partial \Lambda} +\Delta +\Delta^\prime -\Delta_{2,1} -z\frac{\partial}{\partial z}
\right)
\Big]\Psi(z).
\end{align}

  Let us study the constraint equation for $\Psi$ following from the null state condition 
  $((L_{-1})^2 + b^2 L_{-2})\Phi_{2,1} = 0$.
  As analyzed in subsection \ref{subsec:degenerate}, 
  the action of $(L_{-1})^2$ on the degenerate field is simply $(L_{-1})^2 \Phi_{2,1} = \partial^2_z \Phi_{2,1}$.
The action of $L_{-2}$ can be evaluated by extracting the term with the power $w^0$ from (\ref{TPhi21_2}).
The result is
\begin{align}
\label{L-2}
\langle \Delta^\prime , \Lambda^2 |\,L_{-2}\Phi_{2,1}(z) | \Delta,\Lambda^2 \rangle
&=
\Big[ -\frac{1}{z}\frac{\partial}{\partial z}
+\Lambda^2\left(\frac{1}{z} +\frac{1}{z^3} \right)
\nonumber\\
&\qquad\qquad+\frac{1}{2z^2}\left(
\frac{\Lambda}{2} \frac{\partial}{\partial \Lambda} +\Delta +\Delta^\prime -\Delta_{2,1} -z\frac{\partial}{\partial z}
\right)
\Big]\Psi(z).
\end{align}

Now we are ready to complete the formulation of the Schr\"odinger equation.
The null state condition and (\ref{L-2}) imply the following differential equation
\begin{align}
\left[
b^{-2}z^2\frac{\partial^2}{\partial z^2} 
+
\Lambda^2\left(z+\frac{1}{z} \right)
-\frac{3z}{2}\frac{\partial}{\partial z}
+\frac{\Lambda}{4} \frac{\partial}{\partial \Lambda} 
+\frac{\Delta +\Delta^\prime -\Delta_{2,1} }{2}
\right]\Psi(z)=0,
\end{align}
for the irregular conformal block in the presence of the degenerate field.
We will interpret this equation as the Schr\"odinger equation for
the Nekrasov partition function in the presence of the surface operator.
As in subsection \ref{subsec:degenerate}, we recover all $\Omega$-backgrounds $\epsilon_{1,2}$ 
by scaling the parameters as $\Lambda \to \Lambda/\hbar\,,~~ \Delta \to\Delta/\hbar^2$.
Now we take the limit $\epsilon_2\to0$, while keeping $\epsilon_1$ finite.
Then, this limit simplifies the differential equation for the 
normalized function $\Psi^{(0)}(z)= \lim_{\epsilon_2 \rightarrow 0} \Psi(z)\,/\,Z_{\textrm{inst}}$
\begin{align}
\left[
\epsilon_1^2 z^2\frac{\partial^2}{\partial z^2} 
+
\Lambda^2\left(z+\frac{1}{z} \right)
\right]
\Psi^{(0)}(z)
=
\left[a^2
-\frac{\epsilon_1^{2}}{4}
+\frac{\Lambda}{4} \frac{\partial \mathcal{F}(\epsilon_1)}{\partial \Lambda} 
\right]\Psi^{(0)}(z).
\end{align}
This takes the form of the Schr\"odinger equation for the sine-Gordon system
\begin{align}
\left(
-\epsilon_1^2\,\frac{\partial^2}{\partial \theta^2}
+2\Lambda^2 \cos\theta
\right)\Psi(e^{i\theta})
=E\,
\Psi(e^{i\theta}),
\end{align}
where the $\Omega$-background plays the role of the Planck constant.
Notice that the right hand side 
$a^2
+{\Lambda} {\partial_\Lambda \mathcal{F}(\epsilon_1)}/4$
is precisely the classical and the instanton part of $u$.
This quantum Coulomb moduli thus 
corresponds to the energy eigenvalue of the sine-Gordon system.
It is known that the sine-Gordon model is the 2-periodic Toda-chain system.
Since a degeneration of the Hitchin system on a torus with a marked point,
which is the elliptic Calogero-Moser system,
implies the Toda-chain system \cite{Gorsky:1995sr, Marshakov:1997ui},
the sine-Gordon system corresponds to the Hitchin system on a sphere
with two degenerate points.
We expect that a quantum Hitchin system describes the corresponding 
 asymptotically free gauge theory as well as superconformal one.
Then the degeneration of the Hitchin system plays a key role \cite{Nekrasov:1995nq}.

As we explain in appendix \ref{sec:cb}, the wave-function $\Psi$ takes the form 
of the expansion $\sum_{n=0}c_nz^{\delta+n}$.
Let us consider the normalized wave-function
\begin{align}
\psi(z)=z^{-\delta}\Psi(z)
=\exp\left( -\frac{1}{\epsilon_1\epsilon_2}\left(\mathcal{F}(\epsilon_1)
+\epsilon_2\mathcal{W}(\epsilon_1;z) +\cdots\right)\right).
\end{align}
For this normalized correlation function,
the differential equation take the form
\begin{align}
&\Big[
b^{-2}\left(z\frac{\partial}{\partial z}\right)^2 
+b^{-2}\,(2\delta-1)\, z\frac{\partial}{\partial z}
+\delta\left(b^{-2}(\delta-1)-\frac{3}{2} \right)
\nonumber\\
&\qquad\qquad +\Lambda^2\left(z+\frac{1}{z} \right)
-\frac{3z}{2}\frac{\partial}{\partial z}
+\frac{\Lambda}{4} \frac{\partial}{\partial \Lambda} 
+\frac{\Delta +\Delta^\prime -\Delta_{2,1} }{2}
\Big]\psi(z)=0.
\end{align}
Notice that we can eliminate the constant term of this differential operator
by using the following identity
\begin{align}
\delta\left(b^{-2}(\delta-1)-\frac{3}{2} \right)
=
-\frac{\Delta +\Delta^\prime -\Delta_{2,1} }{2}.
\end{align}
Then, the normalized wave-function also satisfies the following simple equation
\begin{align}
\left[
b^{-2}\left(z\frac{\partial}{\partial z}\right)^2 
-2a\,b^{-1}\, z\frac{\partial}{\partial z}
+\Lambda^2\left(z+\frac{1}{z} \right)
+\frac{\Lambda}{4} \frac{\partial}{\partial \Lambda} 
\right]\psi(z)=0,
\end{align}
where we use $b^{-2}(2\delta-1)=-2ab^{-1}+3/2$.
When the $\Omega$-background is recovered by rescaling, we obtain the equation of the form
\begin{align}
\left[
\epsilon_1^2 \left(z\frac{\partial}{\partial z}\right)^2 
-2a\,\epsilon_1\, z\frac{\partial}{\partial z}
+\Lambda^2\left(z+\frac{1}{z} \right)
+\epsilon_1\epsilon_2\frac{\Lambda}{4} \frac{\partial}{\partial \Lambda}
\right]\psi(z)=0.
\end{align}
Note that this equation is not singular at $\epsilon_{1,2}=0$.
Then, we obtain the wave function in the limit $\epsilon_2\to0$
\begin{align}
\left[
\epsilon_1^2\left(z\frac{\partial}{\partial z}\right)^2
+\Lambda^2\left(z+\frac{1}{z} \right)
\right]\psi(z)
=\left(
2a\,z\frac{\partial\mathcal{W}(\epsilon_1;z)}{\partial z}
+\frac{\Lambda}{4} \frac{\partial \mathcal{F}(\epsilon_1)}{\partial \Lambda}\right)
\psi(z).
\end{align}

\subsubsection*{$SU(2)$ gauge theory with one flavor}
Let us next consider the $SU(2)$ gauge theory with one fundamental flavor.
Since the irregular conformal block for the theory is given by 
the inner product of two different Gaiotto states,
the insertion of the degenerate field implies
\begin{align}
\Psi(z)
=\langle \Delta^\prime, \Lambda, m |\Phi_{2,1}(z)|\Delta,{\Lambda^2}/{2}\rangle,
\end{align}
where $m$ corresponds to the mass of the flavor.
We again consider the insertion of the energy-momentum tensor
\bea
\langle \Delta^\prime, \Lambda, m |T(w)\Phi_{2,1}(z)|\Delta,{\Lambda^2}/{2}\rangle
&=&    \left[\sum_{n=0}^\infty w^{-n-2}z^n\left(z\frac{\partial}{\partial z} +\Delta_{2,1}(n+1) \right)
     -\Lambda^2-\frac{2m\Lambda}{w}+\frac{\Lambda^2}{2w^3} \right]\Psi(z)
      \nonumber \\
& &  +\frac{1}{w^2}
\langle \Delta^\prime, \Lambda, m |\Phi_{2,1}(z)L_0|\Delta,{\Lambda^2}/{2}\rangle.\nonumber
\eea
By using $L_0|\Delta,\Lambda, m \rangle=(\Delta+\Lambda\partial_\Lambda)|\Delta,\Lambda, m \rangle$,
we find the following relation:
\begin{align}
\Lambda\frac{\partial \Psi}{\partial \Lambda}=
-(\Delta^\prime+2\Delta)\Psi
+\langle \Delta^\prime, \Lambda, m |[L_0,\Phi_{2,1}(z)]|\Delta,{\Lambda^2}/{2}\rangle
+3\langle \Delta^\prime, \Lambda, m |\Phi_{2,1}(z)L_0|\Delta,{\Lambda^2}/{2}\rangle.\nonumber
\end{align}
Then, we obtain the relation
\begin{align}
&\langle \Delta^\prime, \Lambda, m |T(w)\Phi_{2,1}(z)|\Delta,{\Lambda^2}/{2}\rangle
\nonumber\\
&=\rule{0pt}{4ex}
\left[\frac{z}{w(w-z)}\frac{\partial}{\partial z}
+\frac{\Delta_{2,1}}{(w-z)^2}
-\Lambda^2-\frac{2m\Lambda}{w}+\frac{\Lambda^2}{2w^3}
+\frac{1}{3w^2}\left(
\Lambda\frac{\partial}{\partial \Lambda}
+\Delta^\prime
+2\Delta
-\Delta_{2,1}
-z\frac{\partial}{\partial z}
\right)
\right]\Psi(z).\nonumber
\end{align}
The power expansion in $w$ gives the action of the Virasoro generator $L_{-2}$
\begin{align}
&\langle \Delta^\prime, \Lambda, m |L_{-2}\Phi_{2,1}(z)|\Delta,{\Lambda^2}/{2}\rangle
\nonumber\\
&=\rule{0pt}{4ex}
\left[-\frac{1}{z}\frac{\partial}{\partial z}
-\Lambda^2-\frac{2m\Lambda}{z}+\frac{\Lambda^2}{2z^3}
+\frac{1}{3z^2}\left(
\Lambda\frac{\partial}{\partial \Lambda}
+\Delta^\prime
+2\Delta
-\Delta_{2,1}
-z\frac{\partial}{\partial z}
\right)
\right]\Psi(z).\nonumber
\end{align}
Now we can derive the differential equation for the irregular conformal block. 
The null state condition implies the constraint on $\Psi$
\begin{align}
\left(
b^{-2}z^2\frac{\partial^2}{\partial z^2}
-\frac{4}{3}z\frac{\partial}{\partial z}
+z^2\left(
\frac{\Lambda^2}{2z^3}
-\frac{2m\Lambda}{z}
-\Lambda^2
\right)
+\frac{1}{3}\left(\Lambda\frac{\partial}{\partial \Lambda}
+\Delta^\prime
+2\Delta
-\Delta_{2,1}\right)
\right)\Psi(z)=0.
\end{align}
Let us recover the $\epsilon_{1,2}$ by rescaling the parameters.
We again obtain the Schr\"odinger system in the limit $\epsilon_2\to0$
\begin{align}
\left[
\epsilon_1^{2}z^2\frac{\partial^2}{\partial z^2} 
+z^2\left(
\frac{\Lambda^2}{2z^3}
-\frac{2m\Lambda}{z}
-\Lambda^2
\right)
\right]
\Psi^{(0)}(z)
=
\left[a^2
-\frac{\epsilon_1^{2}}{4}
+\frac{\Lambda}{3} \frac{\partial \mathcal{F}(\epsilon_1)}{\partial \Lambda} 
\right]\Psi^{(0)}(z).
\end{align}
It also takes the form of the Schr\"odinger equation where the potential is similar to $\phi_2^{{\rm SW}}(z)$
in the Seiberg-Witten curve.
The energy eigenvalue is the quantum Coulomb moduli $u(\epsilon_1)$ again.

Let us derive another differential equation for the normalized wave function
$\psi(z)=z^{-\delta}\Psi(z)$.
We can simplify this differential equation
by using the following identities
\begin{align}
&\delta\left(b^{-2}(\delta-1)-\frac{4}{3} \right)
=
-\frac{\Delta^\prime +2\Delta -\Delta_{2,1} }{3},
&b^{-2}(2\delta-1)-\frac{4}{3}=-2ab^{-1}+\frac{1}{6}.
\end{align}
Then, we obtain the differential equation
\begin{align}
\left(
b^{-2}\left(z\frac{\partial}{\partial z}\right)^2
-\left( 2ab^{-1}-\frac{1}{6}\right)z\frac{\partial}{\partial z}
+z^2\left(
\frac{\Lambda^2}{2z^3}
-\frac{2m\Lambda}{z}
-\Lambda^2
\right)
+\frac{1}{3}\Lambda\frac{\partial}{\partial \Lambda}
\right)\psi(z)=0.
\end{align}
Taking the limit $\epsilon_2\to 0$, we obtain the Schr\"odinger equation for the gauge theory with one flavor
\begin{align}
\left(
\epsilon_1^2\left(z\frac{\partial}{\partial z}\right)^2
+z^2\left(
\frac{\Lambda^2}{2z^3}
-\frac{2m\Lambda}{z}
-\Lambda^2
\right)
\right)\psi(z)
=\left(
2a\,z\frac{\partial\mathcal{W}(\epsilon_1;z)}{\partial z}
+\frac{\Lambda}{3} \frac{\partial \mathcal{F}(\epsilon_1)}{\partial \Lambda}\right)
\psi(z).
\end{align}

\subsubsection*{$SU(2)$ gauge theory with two flavors}
Let us move on to the $SU(2)$ gauge theory with two flavors.
The irregular conformal block  with a degenerate field is given by  the inner product 
of the Gaiotto states $|\Delta,\Lambda, m \rangle$ as
\begin{align}
\Psi(z)
=\langle \Delta^\prime, \Lambda, m_2 |\Phi_{2,1}(z)|\Delta,\Lambda, m_1 \rangle.
\end{align}
Again we insert the energy-momentum tensor into the conformal block
\begin{align}
&\langle \Delta^\prime, \Lambda, m_2 |T(w)\Phi_{2,1}(z)|\Delta, \Lambda, m_1 \rangle
\nonumber\\
&\qquad=
\left[\frac{z}{w(w-z)}\frac{\partial}{\partial z}
+\frac{\Delta_{2,1}}{(w-z)^2}
-\Lambda^2-\frac{2m_2\Lambda}{w}-\frac{2m_1\Lambda}{w^3}-\frac{\Lambda^2}{w^4}
\right]\Psi(z)\nonumber\\
&\qquad\qquad\qquad\qquad\qquad\qquad\qquad\quad+\frac{1}{w^2}
\langle \Delta^\prime, \Lambda, m_2 |\Phi_{2,1}(z)L_0|\Delta, \Lambda, m_1 \rangle.
\end{align}
As we have studied in the cases of $N_f=1,2$,
we obtain
\begin{align}
\Lambda\frac{\partial \Psi}{\partial \Lambda}=
-(\Delta^\prime+\Delta)\Psi
+\langle \Delta^\prime, \Lambda, m |[L_0,\Phi_{2,1}(z)]|\Delta,{\Lambda^2}/{2}\rangle
+2\langle \Delta^\prime, \Lambda, m |\Phi_{2,1}(z)L_0|\Delta,{\Lambda^2}/{2}\rangle.
\end{align}
Combining these results, we can write down the expectation value
$\langle \Delta^\prime, \Lambda, m_2 |L_{-2}\Phi_{2,1}(z)|\Delta, \Lambda, m_1 \rangle$.
We then find the differential equation associated with the null state condition:
\begin{align}
\left(
b^{-2}z^2\frac{\partial^2}{\partial z^2}
-\frac{3}{2}z\frac{\partial}{\partial z}
+z^2\left(
-\frac{\Lambda^2}{z^4}
-\frac{2m_1\Lambda}{z^3}
-\frac{2m_2\Lambda}{z}
-\Lambda^2
\right)
+\frac{1}{2}\left(\Lambda\frac{\partial}{\partial \Lambda}
+\Delta^\prime
+\Delta
-\Delta_{2,1}\right)
\right)\Psi(z)=0.\nonumber
\end{align}
In the limit $\epsilon_2\to 0$,
this equation reduces to the following Schr\"odinger equation
\begin{align}
&\left(
\epsilon_1^{2}z^2\frac{\partial^2}{\partial z^2} 
+z^2\left(
-\frac{\Lambda^2}{z^4}
-\frac{2m_1\Lambda}{z^3}
-\frac{2m_2\Lambda}{z}
-\Lambda^2
\right)\right)
\Psi^{(0)}(z)\nonumber\\
&\qquad\qquad\qquad\qquad\qquad\qquad\qquad\qquad\qquad=
\left(a^2
-\frac{\epsilon_1^{2}}{4}
+\frac{\Lambda}{2} \frac{\partial \mathcal{F}(\epsilon_1)}{\partial \Lambda} 
\right)\Psi^{(0)}(z).
\end{align}
Again, the Schr\"odinger equation for the gauge theory with two flavors
has the potential which is similar to $\phi_2^{{\rm SW}}(z)$.
The energy eigenvalue is just the deformed Coulomb moduli $u(\epsilon_1)$.
We expect these characteristics are universal for asymptotically free $\mathcal{N}=2$ gauge theories.

The differential equation for the normalized partition function is also given by
\begin{align}
\left(
b^{-2}\left(z\frac{\partial}{\partial z}\right)^2
- 2ab^{-1}z\frac{\partial}{\partial z}
+z^2\left(
-\frac{\Lambda^2}{z^4}
-\frac{2m_1\Lambda}{z^3}
-\frac{2m_2\Lambda}{z}
-\Lambda^2
\right)
+\frac{\Lambda}{2}\frac{\partial}{\partial \Lambda}\right)
\psi(z)=
0.
\end{align}
By taking the limit $\epsilon_2\to 0$, 
we obtain another Schr\"odinger equation for the gauge theory with $N_f=2$ flavors
\begin{align}
&\left(
(\epsilon_1)^2\left(z\frac{\partial}{\partial z}\right)^2
+z^2\left(
-\frac{\Lambda^2}{z^4}
-\frac{2m_1\Lambda}{z^3}
-\frac{2m_2\Lambda}{z}
-\Lambda^2
\right)
\right)\psi(z)\nonumber\\
&\qquad\qquad\qquad\qquad\qquad\qquad\qquad
=\left(
2a\,z\frac{\partial\mathcal{W}(\epsilon_1;z)}{\partial z}
+\frac{\Lambda}{2} \frac{\partial \mathcal{F}(\epsilon_1)}{\partial \Lambda}\right)
\psi(z).
\end{align}

\subsection{Monodromy of $\Psi$ and relation with quantum integrable system}
\label{subsec:monodromy}
  As found in \cite{AGGTV}, the monodromies of the conformal block 
  with the degenerate field insertion along the $A$- and $B$-cycles correspond 
  to the Wilson and t' Hooft loop operators on the surface operator in the gauge theory.
  In \cite{AGGTV, DGOT}, these monodromies have been calculated in the Liouville theory:
    \bea
    \Psi_\pm (a_i, z + A^j)
    &=&    \exp \left( \mp \frac{2 \pi i a_j}{\epsilon_1} \right) \Psi_\pm(a_i z), 
           \nonumber \\
    \Psi_\pm (a_i, z + B^j)
    &=&    \Psi_\pm (a_i \mp \frac{\epsilon_2}{2} \delta_{ji}, z),
           \label{monodromyPsi}
    \eea
  where $\Psi(z + A({\rm or~}B))$ denotes the monodromy along the $A$(or $B$) cycle.
  The $\pm$ sign in (\ref{monodromyPsi}) reflects the two-fold degeneracy of the solution 
  to the quadratic differential equation obtained in the previous subsections.
  Since $\Psi$ is expanded in $\epsilon_2$ as (\ref{Psiepsilon2}),
  the second equation is equivalent to the condition
    \bea
    \Psi_\pm (a_i, z + B^j)
     =     \exp \left( \mp \frac{1}{2 \epsilon_1} \frac{\partial \CF(\epsilon_1)}{\partial a_j}
         + \CO(\epsilon_2) \right) \Psi_\pm (a_i, z).
    \eea
  These indicate that the monodromies of $\Psi$ around $A$ and $B$ cycles are, in the $\epsilon_2 \rightarrow 0$ limit, 
  the multiplications of the phase factors
  $e^{\mp \frac{2 \pi i a_j}{\epsilon_1}}$ and 
  $e^{\mp \frac{1}{2 \epsilon_1} \frac{\partial \CF(\epsilon_1)}{\partial a_j}}$, respectively.
  Therefore, these lead to the monodromies of $\CW(z; \epsilon_1)$ in (\ref{Psiepsilon2})
    \bea
    \CW_\pm (z + A^j; \epsilon_1)
     =   \pm 2 \pi i a_j, 
           ~~~
    \CW_\pm (z + B^j; \epsilon_1)
     =   \pm \frac{1}{2} \frac{\partial \CF(\epsilon_1)}{\partial a_j}.
           \label{monodromy}
    \eea
  These are reminiscent of the proposal (\ref{MM}).
  We will see that these conditions are indeed related with the analysis in section \ref{sec:quant}.
  Note that the relation between the loop operators in the asymptotically free gauge theory
  and the irregular conformal block with the degenerate field insertion analyzed in the previous subsection
  has not yet been found.
  However, it is natural that the monodromy condition (\ref{monodromyPsi}) holds even in the asymptotically free case.
  
  As we have seen in subsections \ref{subsec:degenerate} and \ref{subsec:irregular}, 
  the differential equation in the limit $\epsilon_2 \rightarrow 0$ becomes generally to
    \bea
    \left( - \epsilon_1^2 \partial^2_z +  V(z; \epsilon_1) \right) \Psi^{(0)}(z)
     =     g(z) u(\epsilon_1) \Psi^{(0)}(z),
    \eea
  where $g(z)$ is a function of $z$ whose choice depends on the choice of a particular gauge theory.
  As observed in the above examples,
  the zero-th order part in $\epsilon_1$ of $V(z; \epsilon_1)$ is the Seiberg-Witten curve modulo moduli dependent term.
  This $V(\epsilon_1)$ is the same one which we have introduced in section \ref{sec:quant} as a potential.
  Furthermore, $u(\epsilon_1) (\propto q \frac{\partial \CF(\epsilon_1)}{\partial q})$ in the right hand side 
  corresponds to the energy $E$ in section \ref{sec:quant}.
  
  Now, recall that $\Psi$ can be written as (\ref{Psiepsilon2}).
  The differential equation is solved order by order as in section \ref{sec:quant}
    \bea
    - x^2 + \epsilon_1 x' + V(z; \epsilon_1)
     =     g(z) u(\epsilon_1).
    \eea
  where we have defined as $\CW = \int^z x(z'; \epsilon_1) dz'$.
  We expand $x$, $V$ and $u$ as
    \bea
    x
     =     \sum_{k=0}^\infty \epsilon_1^k x_k, ~~~
    V
     =     \sum_{k=0}^\infty \epsilon_1^k V_k, ~~~
    u
     =     \sum_{k=0}^\infty \epsilon_1^k u_k.
    \eea
  At lower orders, we obtain
    \bea
    - x_0^2 + V_0
    &=&    g(z) u_0,
           \nonumber \\
    - 2 x_0 x_1 + x'_0 + V_1
    &=&    g(z) u_1,
           \nonumber \\
    - 2 x_0 x_2 - x_1^2 + x'_1 + V_2
    &=&    g(z) u_2, 
    \eea
  and so on.
  Note that compared with the situation in section \ref{sec:quant}, $u(\epsilon_1)$ has $\epsilon_1$-dependence
  which leads to the nonzero values in the right hand sides of higher order equations.
  Similar to (\ref{P0P1P2}), $x_k$ can be written as
    \bea
    x_0
    &=&    \sqrt{V_0 - g(z) u_0}, ~~~
    x_1
     =     \frac{1}{2 x_0} (x'_0 + V_1 - g(z) u_1),
           ~~~
    x_2
     =     \frac{1}{2 x_0} (x'_1 - x_1^2 + V_2 - g(z) u_2).
           \nonumber \\
    \eea
  The $\epsilon_1$-dependence of $u(\epsilon_1)$ has led to the last terms in $x_k$ ($k>0$).
  Note that there could be the choice of sign of $x_0$: $x_0 = \pm \sqrt{V_0 - g u_0}$.
  This would result in the two-fold degeneracy in (\ref{monodromyPsi}).
  Here we have chosen the plus sign for simplicity.
  
  In order to relate this with the proposal in section \ref{sec:quant}, 
  let us consider the contour integral of $xdz$.
  We analyze these only in the lower orders in $\epsilon_1$.
  The contour integral of the differential $x_1 dz$ becomes
    \bea
    \oint x_1 dz
    &=&    \oint \frac{x'_0 + V_1}{2 x_0} dz
         + u_1 \frac{\partial}{\partial u_0} \oint x_0 dz 
     =     \left[ \hat{\CO}_1 + u_1 \frac{\partial}{\partial u_0} \right] \oint x_0 dz,
    \eea
  where $\hat{\CO}_1$ is the same one defined in section \ref{sec:quant}.
  In the next order, the contour integral of $x_2dz$ becomes
    \bea
    \oint x_2 dz
    &=&    \left[ \hat{\CO}_2 + u_1 \frac{\partial}{\partial u_0} \hat{\CO}_1
         + \frac{u_1^2}{2} \frac{\partial^2}{\partial u_0^2} + u_2 \frac{\partial}{\partial u_0} \right] \oint x_0 dz.
    \eea
  Since $\hat{\CO}_1$ and $\hat{\CO}_2$ are the same as the ones considered in section \ref{sec:quant}, 
  the proposal in section \ref{sec:quant} implies that 
    \bea
    \frac{1}{2 \pi i} \oint_A x dz
    &=&    \hat{a}(E=u_0) + (\epsilon_1 u_1 + \epsilon_1^2 u_2 + \ldots) \frac{\partial}{\partial u_0} \hat{a} (E = u_0)
           \nonumber \\
    & &    ~~~~~~~~~~~~
         + \left(\frac{\epsilon_1^2 u_1^2}{2} + \ldots \right) \frac{\partial^2}{\partial u_0^2} \hat{a} (E = u_0)
         + \ldots,
    \eea
  and similar equation for the $B$ cycle integral.
  Therefore, we obtain
    \bea
    \frac{1}{2 \pi i} \oint_A x dz
     =     \hat{a} (E = u(\epsilon_1)), ~~~
    \frac{1}{2} \oint_B x dz
     =     \frac{\partial \hat{\CF}}{\partial \hat{a}} (E = u(\epsilon_1)),
           \label{AB}
    \eea
  up to $\CO(\epsilon_1^3)$ terms.
  At this stage, recall that we have already known the form of $u(\epsilon_1)$, 
  which is the derivative of the deformed prepotential with respect to the gauge coupling constant $\ln q$.
  Recall also that, in section \ref{sec:quant}, we have seen that the following relation holds
    \bea
    u(\epsilon_1; a)
     =     E(\hat{a}; \epsilon_1)|_{\hat{a} \rightarrow a}.
           \label{utoE}
    \eea
  where $E$ is the energy obtained in section \ref{sec:quant} by computing the $A$ cycle integral, 
  namely $E(\hat{a})$ in the step 4.
  It follows from this that the periods (\ref{AB}) are 
    \bea
    \frac{1}{2 \pi i} \oint_A x dz
     =     a, ~~~
    \frac{1}{2} \oint_B x dz
     =     \frac{\partial \CF(\epsilon_1)}{\partial a},
    \eea
  where we have used $\hat{\CF} (\epsilon_1)|_{\hat{a} \rightarrow a} = \CF(\epsilon_1)$.
  These are the expected monodromy conditions satisfied by the conformal block 
  with the degenerate field (\ref{monodromy}).
  
  We have only considered the lower order correction in $\epsilon_1$ above.
  However, we expect that this relation holds for higher orders.
  In summary, if we assume the proposal \cite{MM1} in section \ref{sec:quant} about the deformed prepotential, 
  we recover the expected monodromies of the conformal block.
  Conversely, the monodromy condition (\ref{monodromy}) leads to 
  that the deformed prepotential is indeed obtained by the method in section \ref{sec:quant}.

\section{2d-4d Instantons and Surface Operators}
In this section, we interpret conformal blocks with degenerate field insertion 
in the context of the ramified instanton counting \cite{AGGTV}.
We will focus on the irregular conformal block associated with the pure super Yang-Mills theory for simplicity.

\subsection{Degenerate field insertion and 2d-4d instanton counting}
As we have studied in the previous section, 
the following irregular conformal block would capture the dynamics of the $SU(2)$ super Yang-Mills theory 
with the surface operator 
via the extended AGT conjecture:
\begin{align}
\Psi(
z)=\langle \Delta^{\prime}, \Lambda^2 |\Phi_{2,1}(z)|\Delta, \Lambda^2 \rangle.
\end{align}
Here we take $\Delta^\prime =\Delta(\alpha + b/4)$ and $\Delta =\Delta(\alpha -b/4)$ in accordance with the fusion rule.
In order to give an insight into the instanton counting in the presence of these extended operators,
we will study the irregular conformal block as the Nekrasov partition function.
As we learned from the AGT relation for the pure Yang-Mills theory \cite{Marshakov:2009gn},
the expression in terms of the Shapovalov form
is important for our purpose.
Now let us expand the irregular conformal block by using the formula (\ref{gaiottostate})
\begin{align}
\Psi(z)=\sum_{\vec{Y}}\Lambda^{2|\vec{Y}|}\,
Q_{\Delta}^{-1}([1^{|Y_1|}]; Y_1)\,
Q_{\Delta^{\prime}}^{-1}([1^{|Y_2|}]; Y_2)\,
\langle \Delta^{\prime},Y_2 |\Phi_{2,1}(z)|\Delta, Y_1 \rangle.
\end{align}
In this section, we derive the information of 2d- and 4d-instantons, or ramified instantons, from this expression.

Since we have to expand the irregular conformal block not only in $\Lambda$ but also in  $z$ 
to compare with instanton expansion,
let us expand $\Phi_{2,1}(z)|\Delta, Y \rangle$ in the Verma module $\mathcal{V}_{\Delta^{\prime}}$ as
\begin{align}
\Phi_{2,1}(z)|\Delta, Y \rangle
=\sum_{Y^{\prime}}
z^{|Y^{\prime}|-|Y|+\delta}
\beta^Y_{Y^{\prime}}
\,|\Delta^{\prime}, Y^{\prime} \rangle,
\end{align}
where we define $\delta=\Delta^{\prime}-\Delta_{21}-\Delta$.
See appendix \ref{sec:cb} for details about this formula.
Using this expansion, we can rewrite $\Psi(z)$ as
\begin{align}
\Psi(z)&=\sum_{\vec{Y},Y^{\prime}}\Lambda^{2|\vec{Y}|}\,
z^{|Y^{\prime}|-|Y_1|+\delta}\,
\beta^{Y_1}_{Y^{\prime}}
Q_{\Delta}^{-1}([1^{|Y_1|}]; Y_1)\,
Q_{\Delta^{\prime}}^{-1}([1^{|Y_2|}]; Y_2)\,
Q_{\Delta^{\prime}}(Y_2,\,Y^{\prime} )\nonumber\\
&=z^\delta\,\sum_{n=0}^{\infty}\sum_{{Y}}\Lambda^{2|{Y}|+2n}\,
z^{n-|Y|}
\beta^{Y}_{1^{n}}
Q_{\Delta}^{-1}([1^{|Y|}]; Y).
\end{align}
The point is that we can separate the contributions of 4d- and 2d-instantons as  $\Lambda^{2|{Y}|+2n}\,
z^{n-|Y|}=\Lambda^{4|{Y}|}\,
\lambda^{n-|Y|}$, where $\lambda=\Lambda^2z$ is the 2d-instanton factor.  
Then the 2d-instanton number is counted by $l=n-|Y|$.
This 2d-instanton number can be negative in the presence of the 4d-instanton $k=|Y|\neq0$,
and this configuration represents a 2d-antiinstanton bounded to a 4d-instanton.
See \cite{Gaiotto:2009fs} for related discussion.

\subsection{Explicit computations}
To study the instanton partition function for the surface operator,
we compute the instanton expansion of the normalized partition function $z^{-\delta}\Psi(z)$.
Throughout this section,
we use the formulae for the coefficients $\beta$ which are given in appendix \ref{sec:cb}.

\subsubsection*{$|\textbf{Y}|\textbf{=n=0}$ : constant term}
Let us start with the lowest term.
In our normalization, this term is one as follows:
\begin{align}
\Lambda^0z^0\beta^{\bullet}_{\bullet}\,Q_{\Delta}^{-1}(\bullet\,; \bullet)=1.
\end{align}

\subsubsection*{$|\textbf{Y}|\textbf{=0}$, $\textbf{n=1}$ : one 2d-instanton}
Next we compute the term for $Y=\bullet$ and ${n=1}$.
By using the explicit form of $\beta^{\,\bullet}_{\,1}$ given in appendix \ref{sec:cb}, we find
\bea
\Lambda^2z^1\beta^{\,\bullet}_{\,1}\,Q_{\Delta}^{-1}(\bullet\,; \bullet)
 =\lambda\,\frac{\Delta^\prime+\Delta_{2,1}-\Delta}{2\,\Delta^{\prime}}
 =\lambda\,\frac{b}{2\left(a-\left(\frac{b}{4}+\frac{1}{2b}\right)\right)
} .
\eea
Since this term is proportional to $\lambda=\Lambda^2z$, it is purely one 2d-instanton effect.

\subsubsection*{$|\textbf{Y}|\textbf{=1}$, $\textbf{n=0}$: one 4d-instanton and one 2d-antiinstanton}
The term for $Y=[1]$ and $n=0$ is given by
\bea
\Lambda^2z^{-1}\beta^{1}_{\bullet}\,Q_{\Delta}^{-1}(1; 1)
=- \Lambda^4\lambda^{-1}\frac{\delta}{2\Delta}
=\Lambda^4 \lambda^{-1}\,
\frac{b}{2\left(a+\left(\frac{b}{4}+\frac{1}{2b}\right)\right)} .
\eea
Since the instanton numbers are $(k,l)=(1,-1)$,
the term describes a bound state of a 2d-anti-instanton and a 4d-instanton.

\subsubsection*{$|\textbf{Y}|\textbf{=1}$, $\textbf{n=1}$ : one 4d-instanton}
The term for $Y=[1]$ and $n=1$ describes the purely 4d-instanton contribution:
\begin{align}
\Lambda^4z^{0}\beta^{1}_{1}\,Q_{\Delta}^{-1}(1; 1)
&=
\Lambda^4\left(1-\frac{(1+\delta)(\Delta^\prime+\Delta_{2,1}-\Delta)}{2\Delta^\prime}\right)\frac{1}{2\Delta}\nonumber\\
&=
\Lambda^4\frac{-b^2-2}{4\left(a-\left(\frac{b}{4}+\frac{1}{2b}\right)\right)
\left(a+\left(\frac{b}{4}+\frac{1}{2b}\right)\right)}.
\end{align}

\subsubsection*{$|\textbf{Y}|\textbf{=2}$, $\textbf{n=0}$ : two 4d-instantons and two 2d-antiinstantons}
Let us move on to a little more higher orders.
The following two labels contribute to the instanton for $|Y|=2$ and $n=0$:
\begin{align}
\Lambda^4 z^{-2} \left[\beta^{1^2}_{\bullet}\,Q_{\Delta}^{-1}(1^2; 1^2)
+\beta^{2}_{\bullet}\,Q_{\Delta}^{-1}(1^2; 2)\right]
&=\Lambda^8 \lambda^{-2}\,\left[\delta(\delta-1)\,Q_{\Delta}^{-1}(1^2; 1^2)
+(\Delta_{2,1}-\delta)\,Q_{\Delta}^{-1}(1^2; 2)\right]\nonumber\\
&=\Lambda^8 \lambda^{-2}\,
\frac{b^2}
{8\left(a+\left(\frac{b}{4}+\frac{1}{2b}\right)\right)\left(a+\left(\frac{b}{4}+\frac{1}{b}\right)\right)}.
\end{align}
This term corresponds to the 4d-instanton number $k=2$ and the 2d-instanton number $l=-2$.

\subsubsection*{$|\textbf{Y}|\textbf{=0}$, $\textbf{n=2}$ : two 2d-instantons}
Finally we compute the term for $Y=\bullet$ and $n=2$.
The coefficient $\beta_{\,1^2}$ which is given in appendix \ref{sec:cb} gives the following instanton factor:
\bea
\Lambda^4 z^{2} \beta^{\bullet}_{1^2}\,Q_{\Delta}^{-1}(\bullet; \bullet)
=\lambda^{2}\beta^{\bullet}_{1^2}
=\lambda^{2}\,
\frac{b^2}
{8\left(a-\left(\frac{b}{4}+\frac{1}{2b}\right)\right)\left(a-\left(\frac{b}{4}+\frac{1}{b}\right)\right)}.
\eea
This is the purely two 2d-instanton effect $l=2$.

By combining these results,
we come to the instanton expansion of the partition function 
corresponding to pure $SU(2)$ Yang-Mills theory in the presence of a surface operator:
\begin{align}
\label{2d4dinstanton}
z^{-\delta}\Psi(z)
&=1+\lambda\,\frac{1}{2\,\epsilon_1\left(a-\left(\frac{\epsilon_2}{4}+\frac{\epsilon_1}{2}\right)\right)}
+\Lambda^4 \lambda^{-1}\,
\frac{1}{2\,\epsilon_1\left(a+\left(\frac{\epsilon_2}{4}+\frac{\epsilon_1}{2}\right)\right)}\nonumber\\
&\quad
-\Lambda^4\frac{2\epsilon_1+\epsilon_2}
{4\,\epsilon_1^2\epsilon_2\left(a-\left(\frac{\epsilon_2}{4}+\frac{\epsilon_1}{2}\right)\right)
\left(a+\left(\frac{\epsilon_2}{4}+\frac{\epsilon_1}{2}\right)\right)}
+\Lambda^8 \lambda^{-2}\,
\frac{1}
{8\,\epsilon_1^2\left(a+\left(\frac{\epsilon_2}{4}+\frac{\epsilon_1}{2}\right)\right)\left(a+\left(\frac{\epsilon_2}{4}+\epsilon_1\right)\right)}\nonumber\\
&\quad
+ \lambda^{2}\,
\frac{1}
{8\,\epsilon_1^2\left(a-\left(\frac{\epsilon_2}{4}+\frac{\epsilon_1}{2}\right)\right)\left(a-\left(\frac{\epsilon_2}{4}+\epsilon_1\right)\right)}
+ \cdots.
\end{align}
Here we recover $\epsilon_{1,2}$ by rescaling the parameters. 
In this way we find that
the degenerate field inserted in the irregular conformal block 
describes the Nekrasov-like partition function for ramified instantons.
It supports our expectation that 
we can construct such instanton partition functions by inserting
the degenerate field into conformal block,
without involving mathematics for ramified instantons.
Notice that the result  of our approach 
agrees with the ramified instanton counting \cite{Alday:2010vg}.
In fact, the formula (\ref{2d4dinstanton}) is coincident with
the partition function (B.6) in Appendix.B of \cite{Alday:2010vg} 
through the redefinition $a\to a-\epsilon_2/4$ and the decoupling limit of the adjoint hypermultiplet: 
\begin{align}
m\to \infty,\quad m^2x\to\lambda,\quad m^2y\to\Lambda^4\lambda^{-1}.
\end{align}
The agreement implies a direct relationship between these two different approaches.
This is an important area for further research.

\subsection{Adding fundamental flavors}
It is also straightforward to add matters to the previous results.
For instance,
the degenerate field inserted in the irregular conformal blocks for 
$N_f=1,2$ flavors gives
the following expressions for the corresponding partition functions
\begin{align}
\Psi(z)_{N_f=1}
&=\langle \Delta^\prime, \Lambda, m |\Phi_{2,1}(z)|\Delta,{\Lambda^2}/{2}\rangle\nonumber\\
&=z^\delta\sum_{n,p,q}\sum_{Y}
m^{n-2p}2^{-2q}\Lambda^{n+2q}
z^{n-q}\beta_{\,2^p\cdot 1^{n-2q}}^{Y}Q_\Delta^{-1}(1^q,Y).
\\
\Psi(z)_{N_f=2}
&=\langle \Delta^\prime, \Lambda, m_2 |\Phi_{2,1}(z)|\Delta,\Lambda, m_1\rangle\nonumber\\
&=z^\delta\sum_{n_{1,2}\,,p_{1,2}}\sum_{Y}
m_1^{n_1-2p_1}m_2^{n_2-2p_2}\Lambda^{n_1+n_2}
z^{n_1-n_2}\beta_{\,2^p_1\cdot 1^{n_1-2p_2}}^{Y}Q_\Delta^{-1}(2^{p_2}\cdot 1^{n_2-2p_2},Y).
\end{align}    
It would be interesting to study the structure of these correlators 
and rewrite it as Nekrasov-like partition functions.

\section{Conclusion and Discussion}
\label{sec:conclusion}
  In this paper, we have considered the relation 
  between $\CN=2$ supersymmetric gauge theories and quantum integrable systems.
  We have seen that the deformed prepotential can be obtained from the monodromies of the wave-function
  which is calculated from the Schr\"odinger equation of the integrable system.
  We have then derived this equation from the conformal block with the degenerate field insertion.
  By using the AGT relation, we have successfully related the deformed prepotential emerging from the wave-function
  with monodromy operation of the conformal block with the degenerate field.
  We have also studied the instanton counting of the instanton partition function with the surface operator 
  which corresponds to the irregular conformal blocks with a degenerate field.
  
  In this paper, we concentrated on the case with the $SU(2)$ gauge group which corresponds to the Liouville theory.
  It would be important to consider higher rank generalization.
  In \cite{MM2, Popolitov}, the proposal \cite{MM1} in section \ref{sec:quant} has been checked 
  for the $SU(N)$ pure super Yang-Mills theory by analyzing the corresponding Baxter equation.
  It would be interesting to consider such the differential equation in the point of view of the Toda field theory.
  The loop operators in the Toda theory \cite{DGG, Passerini} might be related with the analysis in \cite{MM2}.
  
  The correspondence between the Hitchin systems and Nekrasov-Shatashvili's integrable systems
  also merits intensive investigation.
  Our result suggests that Hitchin systems of degenerated type
  are associated with asymptotically free gauge theories.
  It is therefore important to study the degenerated Hitchin systems from the perspective of the AGT relation
  \cite{Nanopoulos:2009uw, Nanopoulos:2010zb, Nanopoulos:2010ga}.
  The analysis of the Hitchin system from M-theory perspective \cite{BT} would be useful.
  
  In section 4, we recast the degenerate irregular conformal block into the Nekrasov-like partition function.
  The further study of these conformal blocks would give us a fresh insight into the instanton counting
  in the presence of surface operators.
  For higher rank theories, this formulation in terms of the Virasoro algebra
  should be extended for the $\mathcal{W}$-algebra  \cite{Wyllard, MM, Taki:2009zd}.
  
  In \cite{Kozcaz:2010af}, the matrix model description for surface operators
  was given for the case of $\epsilon_1+\epsilon_2=0$.
  It would be interesting to study the monodromic characteristic of the wave-function by using the matrix models.
  The matrix model for higher rank theories \cite{DV, IMO, Schiappa:2009cc}
  would help us to study surface operators of $SU(N)$ gauge theories.
  
  Extended observables such as the Wilson loops are important to probe the phase structure of
  gauge theories.
  The richness of phases of $\mathcal{N}=1$ gauge theories is well-known \cite{Cachazo:2002zk},
  and there are many phases which we cannot distinguish by the Wilson-'t Hooft operators.
  $\mathcal{N}=1$ analogues of surface operators would play an important role to classify these phases,
  and we expect that it will work for the $\CN=1$ version of Gaiotto quivers \cite{MTTY, Benini:2009mz}.

\section*{Acknowledgements}
  We would like to thank Giulio Bonelli, Tohru Eguchi, Kazuo Hosomichi, Hiroshi Itoyama and Alessandro Tanzini 
  for useful comments.
  K.M. would like to thank KIAS and SISSA for warm hospitality during part of this project.
  He also would like to thank the organizers of the conference ``Recent Advances in Gauge Theories and CFTs"
  at Yukawa Institute for Theoretical Physics, Kyoto University (1 -- 2 March 2010).
  Research of K.M. is supported in part by JSPS Bilateral Joint Projects (JSPS-RFBR collaboration).
  M.T. is supported by JSPS Grant-in-Aid for Creative Scientific Research, No.19GS0219.

\appendix

\section*{Appendix}

\section{Nekrasov's Instanton Partition Function}
\label{sec:Nekrasov}

The Nekrasov partition function is a generating function of the Seiberg-Witten prepotential.
Formally, the partition function is defined by the regularized volume of the instanton moduli space
\begin{align}
\label{Nekrasov}
Z_{\textrm{inst}}(\vec{a},\Lambda,\epsilon_1,\epsilon_2)=\sum_{k=0}^{\infty}q^k
\int_{\mathcal{M}_{N_c,k}}
d \,\textrm{Vol}_{\,\,\vec{a},\epsilon_1,\epsilon_2},
\end{align}
where $d \,\textrm{Vol}_{\,\,\vec{a},\epsilon_1,\epsilon_2}$ is an instanton measure factor
with an equivariant torus action.
We can obtain the precise form of it
by applying the equivariant localization method to the path integral over the instanton moduli space.
For instance the partition function for  $SU(N)$ supersymmetric gauge theory with hypermultiplets takes the form
\cite{Nekrasov, Flume:2002az, Bruzzo:2002xf}
\begin{align}
\label{Nekrasov}
Z_{\textrm{inst}}(\vec{a},\Lambda,\epsilon_1,\epsilon_2)
=\sum_{\vec{Y}}\frac{q^{ |\vec{Y}|}}
{\prod_{\alpha,\beta=1}^{N_c}n_{\alpha,\beta}^{\vec{Y}}(\vec{a},\epsilon_1,\epsilon_2)}\,
z_{\textrm{matters}}(\vec{a},\vec{Y},\vec{m},\epsilon_1,\epsilon_2),
\end{align}
where $\vec{Y}=(Y_1,\cdots,\,Y_{N_c})$ is a vector consists of $N_c$ Young diagrams, 
and its norm $|\vec{Y}|$ is defined by $\sum_n |Y_n|$.
Here $\vec{a}$ and $\epsilon_{1,2}$ are the weights of the maximal torus action 
$U(1)^{N_c-1} \times U(1) \times U(1)$,
which is the Cartan of the isometry of the instanton moduli space $SU(N_c)\times SU(2)^2\simeq SU(N_c)\times SO(4)_L$.
The expansion factor $q$ is the dynamical scale $\Lambda^{2N_c-N_f }$ for gauge theory with $N_f<2N_c$ flavors, 
but, on the other hand, it is the UV gauge coupling constant $q=e^{2\pi i\tau}$ for the superconformal theory $N_f=2N_c$.

The partition function (\ref{Nekrasov}) consists of the contributions of the vector multiplet and the hypermultiplets.
The denominator $\prod n_{\alpha,\beta}^{\vec{Y}}$ comes of the vector multiplet integral.
The precise form is given by the eigenvalues of the torus action 
on the tangent space of the moduli space
\begin{align}
\label{denominator}
n_{\alpha,\beta}^{\vec{Y}}(\vec{a},\epsilon_1,\epsilon_2)
&=\prod_{(i,j)\in Y_{\alpha}} 
(a_{\alpha}-a_{\beta} -l_{Y_{\beta}}(i,j)\epsilon_1+(a_{Y_{\alpha}}(i,j)+1)\epsilon_2)\nonumber\\
&\qquad\times 
\prod_{(i,j)\in Y_{\beta}} 
(a_{\alpha}-a_{\beta} +(l_{Y_{\alpha}}(i,j)+1)\epsilon_1-a_{Y_{\beta}}(i,j)\epsilon_2 ).
\end{align}
$\vec{a}=(a_1,\cdots,\, a_N)$ is the eigenvalues of the adjoint scalar field.
An arm length and leg length of a Young diagram are defined by $a_Y(i,j)=Y_i-j$ and $l_Y(i,j)={Y^t}_j-i$.

The contributions of matter fields come of the matter bundle over the instanton moduli space, 
which is the bundle of the Dirac zero modes in the representation of the matter field we are interested in.
Since these zero modes are fermionic, this contribution appears, 
when we apply the localization method to the path integral,
in the numerator of the instanton measure.
For (anti)fundamental hypermultiplet, the instanton measure factor is
\begin{align}
\label{numeratorfund}
&z_{\textrm{fund.}}(\vec{a},\vec{Y},m\,; \epsilon_1,\epsilon_2)
=\prod_{\alpha=1}^{N_c}\prod_{(i,j)\in Y_{\alpha}} 
(a_{\alpha}+\epsilon_1(i-1)+\epsilon_2 (j-1)-m+\epsilon),\\
\label{numeratorantifund}
&z_{\textrm{antifund.}}(\vec{a},\vec{Y},m\,; \epsilon_1,\epsilon_2)
=z_{\textrm{fund.}}(\vec{a},\vec{Y},\epsilon-m; \epsilon_1,\epsilon_2).
\end{align}
The adjoint matter bundle is the tangent bundle of the instanton moduli space.
It contributes to the instanton measure as
\begin{align}
\label{numeratoradj}
z_{\textrm{adj.}}(\vec{a},\vec{Y},m; \epsilon_1,\epsilon_2)
&=\prod_{\alpha,\beta=1}^{N_c}\prod_{(i,j)\in Y_{\alpha}} 
(a_{\alpha}-a_{\beta}-l_{Y_{\beta}}(i,j)\epsilon_1+(a_{Y_{\alpha}}(i,j)+1)\epsilon_2-m )\nonumber\\
&\qquad\qquad\times 
\prod_{(i,j)\in Y_{\beta}} 
(a_{\alpha}-a_{\beta}+(l_{Y_{\alpha}}(i,j)+1)\epsilon_1-a_{Y_{\beta}}(i,j)\epsilon_2-m ).
\end{align}
Notice that $z_{\textrm{vec.}}(\vec{Y})^{-1}=\prod n_{\alpha,\beta}^{\vec{Y}}=
z_{\textrm{adj.}}(\vec{Y},m=0)$
since the vector multiplet also transforms in the adjoint representation
and this multiplet gives a bosonic contribution.

Let $Z_k$ be the $k$-instanton part of the partition function:
\begin{align}
Z_{\textrm{inst}}(\vec{a},\Lambda,\epsilon_1,\epsilon_2)
=\sum_{k=0}^\infty{q^k}\,Z_{\,k}(\vec{a},\epsilon_1,\epsilon_2).
\end{align}
We  compute $1$ and $2$-instanton partition functions theories in what follows.

\subsection*{1-instanton}
Terms with $|\vec {Y}|=1$ contribute to $1$-instanton part of the Nekrasov partition function (\ref{Nekrasov}).
Such Young diagrams take the form of $\vec{Y}=(\,\fundl\,,\bullet,\bullet,\cdots),\, (\bullet, \fundl\,,\bullet,\cdots),\, \cdots$.
For adjoint hypermultiplets, the contribution of the fixed point $\vec{Y}=(\fundl\,,\bullet,\bullet,\cdots)$ to the instanton measure is given by
\begin{align}
&z_{\textrm{adj.}}(\vec{a},(\fundl\,,\bullet,\bullet,\cdots),m; \epsilon_1,\epsilon_2)\nonumber\\
&=(\epsilon_1-m)(\epsilon_2-m )\prod_{\beta\neq1}^{N_c}
(a_{1}-a_{\beta}+\epsilon-m )(-a_1+a_{\beta}-m ).
\end{align}
The vector multiplet factor is  $z_{\textrm{vec.}}(\vec{Y})=
1/z_{\textrm{adj.}}(\vec{Y},m=0)$.
For fundamental hypermultiplets, their contribution is
\begin{align}
z_{\textrm{fund.}}(\vec{a},(\fundl\,,\bullet,\bullet,\cdots),m; \epsilon_1,\epsilon_2)
&=
(a_{1}-m+\epsilon),\\
z_{\textrm{antifund.}}(\vec{a},(\fundl\,,\bullet,\bullet,\cdots),m; \epsilon_1,\epsilon_2)
&=z_{\textrm{fund.}}(\vec{a},(\fundl\,,\bullet,\bullet,\cdots),\epsilon-m; \epsilon_1,\epsilon_2)\nonumber\\
&=(a_{1}+m).
\end{align}
Thus $1$-instanton Nekrasov partition functions for $SU(N_c)$ gauge theory with $N_f$ fundamentals or an adjoint:
\begin{align}
\label{Nekrasovfund}
&Z_{\,N_f,\,k=1}(\vec{a},m,\epsilon_1,\epsilon_2)
=\sum_{\alpha=1}^{N_c} \frac{\prod_{f=1}^{N_f}(a_{\alpha}-m_f+\epsilon)}
{\epsilon_1\epsilon_2   \prod_{\beta( \neq \alpha)}^{N_c}  a_{\beta,\alpha} ( a_{\alpha,\beta}+\epsilon)},
\\
\label{Nekrasovadj}
&Z_{\,\mathcal{N}=2^*,\,k=1}(\vec{a},m,\epsilon_1,\epsilon_2)
=\sum_{\alpha=1}^{N_c} \frac{(\epsilon_1-m)(\epsilon_2-m)}
{\epsilon_1\epsilon_2 }
 \prod_{\beta( \neq \alpha)}^{N_c}
\frac{(a_{\alpha ,\beta}+\epsilon-m)(a_{\alpha ,\beta}+m)}
{  a_{\alpha,\beta} ( a_{\alpha,\beta}+\epsilon)}.
\end{align}
For $SU(2)$ gauge theory with an adjoint,
the 1-instanton partition function take the form:
\begin{align}
Z_{\,\mathcal{N}=2^*,\,k=1}(\vec{a},m,\epsilon_1,\epsilon_2)
=
-2\,{\frac { \left( -\epsilon _{{1}}+m \right) 
 \left( \epsilon_{{2}}-m \right)  \left( 4\,{a}^{2}-{\epsilon _{{1}}}^
{2}-2\,\epsilon _{{1}}\epsilon_{{2}}+\epsilon _{{1}}m-{\epsilon_{{2}}}
^{2}+\epsilon_{{2}}m-{m}^{2} \right) }{\epsilon _{{1}}\epsilon_{{2}}
 \left( 2\,a+\epsilon _{{1}}+\epsilon_{{2}} \right)  \left( 2\,a-
\epsilon _{{1}}-\epsilon_{{2}} \right) }}.
\end{align}

\subsection*{2-instanton}
Let us consider $N_c=2$ gauge theories for simplicity.
The Young diagrams which contribute to the 2-instanton partition function must satisfy $|\vec{Y}|=2$.
There are three types of such Young diagrams:
$\vec{Y}=(\,\fundl\,,\fundl\,)$, $(\anti\,,\bullet\,)$, $(\,\symm,\bullet\,) \cdots$.
Let us compute the contribution of $\vec{Y}=(\,\fundl\,,\fundl\,)$.
The adjoint factor becomes
\begin{align}
&z_{\textrm{adj.}}   (\vec{a},(\,\fundl\,,\,\fundl\,) ,m; \epsilon_1,\epsilon_2) 
=( \epsilon_1-m)^2( \epsilon_2-m)^2\nonumber\\
&\qquad\qquad \times
( a_{12}+\epsilon_1-m)( a_{12}-\epsilon_1+m)( a_{12}+\epsilon_2-m)( a_{12}-\epsilon_2+m).
\end{align}
The contribution of the fixed point $\vec{Y}=(\,\anti\,,\bullet\,)$ is also given by the following polynomial:
\begin{align}
&z_{\textrm{adj.}}   (\vec{a},(\,\anti\,,\bullet\,) ,m; \epsilon_1,\epsilon_2) 
=( \epsilon_1-m)( \epsilon_2-m)( \epsilon_1-\epsilon_2-m)( 2\epsilon_2-m)\nonumber\\
&\qquad\qquad\qquad \times
( a_{12}+m)( a_{12}+\epsilon_1+\epsilon_2-m)(a_{12}+ \epsilon_2+m)(a_{12}+\epsilon_1+2\epsilon_2-m).
\end{align}
Finally, the Young diagram $\vec{Y}=(\,\symm,\bullet\,)$ gives
\begin{align}
&z_{\textrm{adj.}}   (\vec{a},(\,\symm,\bullet\,) ,m; \epsilon_1,\epsilon_2) 
=( \epsilon_1-m)( \epsilon_2-m)( -\epsilon_1+\epsilon_2-m)( 2\epsilon_1-m)\nonumber\\
&\qquad\qquad\qquad \times
( a_{12}+m)( a_{12}+\epsilon_1+\epsilon_2-m)(a_{12}+ \epsilon_1+m)(a_{12}+2\epsilon_1+\epsilon_2-m).
\end{align}
Notice that $z_{\textrm{adj.}}   (\,a_1,a_2,(\,Y_1,Y_2\,))=z_{\textrm{adj.}}   (\,a_2,a_1,(\,Y_2,Y_1\,))$.
For instance, the $2$-instanton partition function for $SU(2)$ gauge theory with an adjoint is
\begin{align}
\label{Nekrasov2}
&Z_{\,\mathcal{N}=2^*,\,k=2}(\vec{a},\epsilon_1,\epsilon_2)
\rule{0pt}{5ex}=
\frac{z_{\textrm{adj.}}   (\vec{a},(\,\fundl\,,\,\fundl\,) ,m; \epsilon_1,\epsilon_2) }
{z_{\textrm{adj.}}   (\vec{a},(\,\fundl\,,\,\fundl\,) ,0; \epsilon_1,\epsilon_2) }
+\frac{z_{\textrm{adj.}}   (\vec{a},(\,\anti\,,\bullet\,) ,m; \epsilon_1,\epsilon_2)}
{z_{\textrm{adj.}}   (\vec{a},(\,\anti\,,\bullet\,) ,0; \epsilon_1,\epsilon_2)}
\nonumber\\
 &\rule{0pt}{5ex} 
 +\frac{z_{\textrm{adj.}}   (-\vec{a},(\,\anti\,,\bullet\,) ,m; \epsilon_1,\epsilon_2)}
{z_{\textrm{adj.}}   (-\vec{a},(\,\anti\,,\bullet\,) ,0; \epsilon_1,\epsilon_2)}
 +
 \frac{z_{\textrm{adj.}}   (\vec{a},(\,\symm,\bullet\,) ,m; \epsilon_1,\epsilon_2)}
 {z_{\textrm{adj.}}   (\vec{a},(\,\symm,\bullet\,) ,0; \epsilon_1,\epsilon_2)}
+ \frac{z_{\textrm{adj.}}   (-\vec{a},(\,\symm,\bullet\,) ,m; \epsilon_1,\epsilon_2)}
 {z_{\textrm{adj.}}   (-\vec{a},(\,\symm,\bullet\,) ,0; \epsilon_1,\epsilon_2)}\nonumber\\
 &\rule{0pt}{5ex}
 =-{\frac { \left( 8\,{m}^{6}{\epsilon_{{1}}}^{2}+8\,{m}^{6}{\epsilon _{
{2}}}^{2}-128\,{a}^{6}{m}^{2}-8\,{m}^{6}{a}^{2}+\cdots+20\,{\epsilon _{{1}}}^
{7}\epsilon_{{2}}+20\,\epsilon _{{1}}{\epsilon_{{2}}}^{7} \right) 
 }{
 {\epsilon _1}^2{\epsilon_2}^{2}
 \left( 2\,a-\epsilon_{{1}}-2\,\epsilon _{{2}} \right)  \left( 2\,a+
\epsilon_{{1}}+2\,\epsilon _{{2}} \right)  \left( 2\,a-\epsilon _{{2}}
-2\,\epsilon_{{1}} \right) 
  \left( 2\,a+\epsilon _{{2}}+2\,\epsilon_{{1}} \right) 
} }
\nonumber\\&\quad\rule{0pt}{5ex}
\times
\frac{ \left( \epsilon _{{1}}-m \right)  \left( \epsilon _{{2}}-m \right)}
{ \left( 2\,a-\epsilon _{{1}}-\epsilon_{{2}}
 \right) \left( 2\,a+\epsilon _{{1}}+\epsilon_{{2}} \right) }
 .
\end{align}
We can compute the deformed prepotential by using the above results as
\begin{align}
-\frac{1}{\epsilon_1\epsilon_2}\mathcal{F}_{\textrm{inst}}
&=\log\left(1+qZ_{k=1}+q^2Z_{k=2}+\cdots \right)\nonumber\\
&=qZ_{k=1}+q^2\left(Z_{k=2}-\frac{1}{2}Z_{k=1}^2 \right)+\cdots.
\end{align}

\subsection*{classical and perturbative part}
The classical part of the partition function is given by
\begin{align}
Z_{\textrm{class}}=
\exp \left(-\frac{2\pi i}{\epsilon_1\epsilon_2}\tau a^2\right).
\end{align}
This part corresponds to the
gauge coupling term of the action
$\partial^2{\mathcal{F}_{\textrm{class}}}\propto \tau$.

The perturbative parts are given by Barne's double-Gamma function
\bea
\Gamma_2(x|\epsilon_1,\epsilon_2)
&=&
\exp \gamma_{\epsilon_1,\epsilon_2}(x-\epsilon)
=\exp \frac{d}{ds}\left[\frac{1}{\Gamma(s)}\int_0^\infty 
\frac{t^{s-1}\,e^{-tx}\,dt}{(1-e^{-\epsilon_1t})(1-e^{-\epsilon_2t})}\right]_{s=0}
\nonumber\\
&\sim& \prod_{m,n=0}^\infty \frac{1}{x+m\epsilon_1+n\epsilon_2}.
\eea
See \cite{Nekrasov:2003rj} for details of the function.
The perturbative instanton measures are then given by
\begin{align}
&z_{\textrm{vect}}^{\textrm{pert}}
=\prod_{i<j}\,\Gamma_2(a_{ij}+\epsilon_1|\epsilon_1,\epsilon_2)^{-1}
\Gamma_2(a_{ij}+\epsilon_2|\epsilon_1,\epsilon_2)^{-1},\\
&z_{\textrm{fund}}^{\textrm{pert}}
=\prod_{i}\,
\Gamma_2(a_{i}+\epsilon-m|\epsilon_1,\epsilon_2),\\
&z_{\textrm{adj}}^{\textrm{pert}}
=\prod_{i,j}\,\Gamma_2(a_{ij}+\epsilon-m|\epsilon_1,\epsilon_2),
\end{align}
where we follows the convention of \cite{AGT}.

\section{Calculation of $E$}
\label{sec:ellipticintegral}
  In this appendix, we calculate the energy at zero-th order in $\epsilon_1$ 
  which was used in the analysis in section \ref{sec:quant}.
  In the gauge theory point of view, this corresponds to the evaluation of the Coulomb moduli $u$.
  We consider the models corresponding to the $\CN=2^*$ gauge theory 
  and the $SU(2)$ gauge theory with four flavors in turn.

\subsubsection*{$\CN=2^*$ $SU(2)$ gauge theory}
  First of all, we fix our notation.
  The Weierstrass elliptic function $\CP$ is double periodic with periods $\pi$ and $\pi \tau$
  and is expressed as
    \bea
    \CP(z)
    &=&  - \zeta'(z), ~~~
    \zeta(z)
     =     \frac{\vartheta'_1(z|\tau)}{\vartheta_1(z|\tau)} + 2 \eta_1 z, 
           \nonumber \\
    \eta_1
    &=&  - \frac{2 \pi i}{3} 
           \frac{\frac{\partial}{\partial \tau} \vartheta'_1(z|\tau)|_{z=0}}{\vartheta'_1(z|\tau)|_{z=0}}
     =   - \frac{1}{6} \frac{\vartheta'''_1(z|\tau)|_{z=0}}{\vartheta'_1(z|\tau)|_{z=0}},
    \eea
  where $\vartheta_1(z|\tau)$ is elliptic theta function.
  The Weierstrass function satisfies
    \bea
    \CP(z)'
     =     4 \CP(z)^3 - g_2 \CP(z) - g_3,
    \eea
  where 
    \bea
    g_2
    &=&    \frac{4}{3} 
           \left( 1 + 240 \sum_{n=1}^\infty \frac{n^3 q^{n}}{1 - q^{n}} \right),
           \nonumber \\
    g_3
    &=&    \frac{8}{27} 
           \left( 1 - 504 \sum_{n=1}^\infty \frac{n^5 q^{n}}{1- q^{n}} \right),
           \label{g2g3}
    \eea
  We also define $g_1 = - 2 \eta_1$ whose expansion is 
    \bea
    g_1
     =   - \frac{1}{3} 
           \left( 1 - 24 \sum_{n=1}^\infty \frac{n q^{n}}{1- q^{n}} \right).
           \label{g1}
    \eea
  
  We consider $A$ cycle integral of $P_0$ (\ref{EN=2*})
    \bea
    \pi 
     =     \oint_A \sqrt{\CE - \frac{M}{4}\CP(z)},
    \eea
  where $\CE$ are expanded as $\CE = 1 + M \CE_1(q) + M^2 \CE_2(q) + \ldots$.
  (The coefficients $\CF_k(q)$ are functions only of $q$.)
  These coefficients can be written in terms of $f_n$ defined by
    \bea
    f_n
     =     \frac{1}{\pi} \oint_A \CP(z)^n dz,
    \eea
  as
    \bea
    \CE_1(q)
    &=&    \frac{f_1}{4}, ~~~
    \CE_2(q)
     =     \frac{\CE_1(q)^2}{4} - \frac{\CE_1(q) f_1}{8} + \frac{f_2}{64}, ~~~
           \nonumber \\
    \CE_3(q)
    &=&    \frac{\CE_2 \CE_1}{2} - \frac{\CE_2 f_1}{8} - \frac{\CE_1^3}{8} + \frac{3 \CE_1^2 f_1}{32}
         - \frac{3 \CE_1 f_2}{128} + \frac{f_3}{512}.
    \eea
  Since $f_n$ are written in terms of $g_i$ as 
    \bea
    f_1
     =     g_1, ~~~
    f_2
     =     \frac{g_2}{12}, ~~~
    f_3
     =     \frac{g_3}{10} + \frac{3 g_1 g_2}{20},
    \eea
  the coefficients are expressed as
    \bea
    \CE_1(q)
    &=&    \frac{g_1}{4}
     =   - \frac{1}{12} + 2 q + 6 q^2 + 8 q^3 + \ldots,
           \nonumber \\
    \CE_2(q)
    &=&    \frac{g_2}{768} - \frac{g_1^2}{64}
     =     \frac{1}{2} q + 3 q^2 + 6 q^3 + \ldots,
           \nonumber \\
    \CE_3(q)
    &=&    \frac{g_3 - g_1 g_2}{5120} + \frac{g_1^3}{256}
     =   - \frac{3}{2} q^2 - 12 q^3 + \ldots,
    \eea
  Therefore, the energy $E(a) \equiv 4 a^2 \CE$ can be written as
    \bea
    E(a)
    &=&    4 \left( a^2 - \frac{m^2}{12} + \frac{m^2 (4 a^2 + m^2)}{2 a^2}q 
         + \frac{m^2 (192 a^6 + 96 m^2 a^4 - 48 m^4 a^2 + 5 m^6)}{32 a^6}q^2 + \ldots \right)
           \nonumber \\
    \eea

\subsubsection*{$SU(2)$ gauge theory with four flavors}
  The zero-th order one-form $P_0 dz$ (\ref{P0P1}) can be written as
    \bea
    P_0
     =     \frac{\sqrt{P_4(z)}}{z(z-1)(z-q)},
    \eea
  where $P_4$ is the following polynomial of degree $4$:
    \bea
    P_4
    &=&    \tilde{m}_0^2 z^4 + \left( - (1+2q) \tilde{m}_0^2 - \tilde{m}_1^2 + m_0^2 + (2q-1)m_1^2 + (q-1)E \right) z^3
           \nonumber \\
    & &  + \left( q (q+2)\tilde{m}_0^2 + (1 + 2q) \tilde{m}_1^2 - 2q m_0^2 
         - (q^2 + 2q - 1) m_1^2 + (1 - q^2) E \right) z^2
           \nonumber \\
    & &  + \left( -q^2 \tilde{m}_0^2 - q (q + 2) \tilde{m}_1^2 + q^2 (m_0^2 + m_1^2) + q (q-1)E \right) z
         + \tilde{m}_1^2 q^2.
    \eea
  Also, the derivative with respect to $E$ defines the holomorphic one-form:
    \bea
    \omega_0
     =     \frac{\partial (P_0 dz)}{\partial E} 
     =     \frac{(q-1)dz}{\sqrt{P_4(z)}}.
    \eea
  
  For simplicity, we consider the equal hypermultiplet mass case.
  This implies that $\tilde{m}_0 = \tilde{m}_1 = 0$ and $m_0 = m_1 = m$.
  Note that these parameters are related with the hypermultiplets masses $\mu_i$ as
    \bea
    m_0
    &=&    \frac{1}{2} (\mu_1 + \mu_2), ~~~
    m_1
     =     \frac{1}{2} (\mu_3 + \mu_4),
           \nonumber \\
    \tilde{m}_0
    &=&    \frac{1}{2} (\mu_1 - \mu_2), ~~~
    \tilde{m}_1
     =     \frac{1}{2} (\mu_3 - \mu_4).
           \label{massSU(2)}
    \eea
  In this choice of the masses, by redefining $E = \tilde{E} - \frac{2q m^2}{(q-1)}$, the polynomial reduces to
  the degree $3$ polynomial:
    \bea
    P_3(z)
     =     (q-1) \tilde{E} z (z - z_+) (z - z_-),
    \eea
  with 
    \bea
    z_\pm
     =     \frac{1}{2} \left( 1 + q + (1-q) \frac{m^2}{\tilde{E}} 
         \pm (1-q) \sqrt{1 + \frac{2(1+q)}{1-q} \frac{m^2}{\tilde{E}} + \frac{m^4}{\tilde{E}^2}} \right).
    \eea
  In this case, the holomorphic one-form becomes
    \bea
    \omega_0
     =     \frac{(q-1)dz}{\sqrt{P_3(z)}}
     =     \frac{1}{2} \sqrt{\frac{q-1}{z_+ \tilde{E}}} \frac{dz}{\sqrt{z (1-z)(1- k^2z)}},
    \eea
  where in the last equality we have rescaled $z$ as $z \rightarrow z z_-$ and $k^2 = z_-/z_+$.
  
  In order to obtain the expression for the energy in terms of $a$, 
  we take a derivative of $a = \frac{1}{2 \pi i } \oint_A P_0 dz$ with respect to $\tilde{E}$
    \bea
    \frac{\partial a}{\partial \tilde{E}}
     =     \frac{1}{2 \pi i } \oint_A \omega_0
     =     \frac{1}{2} \sqrt{\frac{1-q}{z_+ \tilde{E}}} F(\frac{1}{2}, \frac{1}{2}, 1; k^2),
    \eea
  where $F(a,b,c; k^2)$ is the hypergeometric function.
  We expand the right hand side in the large $E$ region 
  as $\sqrt{\frac{1-q}{z_+ \tilde{E}}} F(\frac{1}{2}, \frac{1}{2}, 1; k^2)
   = (E)^{-1/2} (h_0(q) + h_1(q) \frac{m^2}{\tilde{E}} + h_2(q) \frac{m^4}{\tilde{E}^2} + \ldots)$
  where $h_i(q)$ are functions of only $q$.
  After integrating by $\tilde{E}$, we obtain
    \bea
    a 
     =     \sqrt{\tilde{E}} \left( h_0(q) - h_1(q) \frac{m^2}{\tilde{E}} - \frac{h_2(q)}{3} \frac{m^4}{\tilde{E}^2}
         + \ldots \right).
    \eea
  We then solve this in terms of $\tilde{E}$:
    \bea
    \tilde{E}
     =     \frac{a^2}{h_0^2} \left( 1 + 2 h_0 h_1 \frac{m^2}{a^2} + \frac{h_0^2 (2 h_0 h_2 - 3 h_1^2)}{3} \frac{m^4}{a^4}
         + \ldots \right)
    \eea
  Finally, by returning to the original $E = \tilde{E} - \frac{2q m^2}{(q-1)}$, we obtain
    \bea
    E
    &=&    a^2 - m^2 + \frac{a^4 + 2 m^2 a^2 + m^4}{2a^2}q
         + \frac{13a^8 + 36m^2 a^6 + 22m^4 a^4 - 12m^6 a^2 + 5m^8}{32a^6} q^2 + \CO(q^3).
           \nonumber \\
    \eea

\section{Action of Degenerate Field on Verma Module}
\label{sec:cb}
In this appendix,
we provide the quantities we use in section 4 to compute the instanton partition function via the AGT relation.
\subsection{Kac determinant}
In section 4, we report the instanton partition function in the presence of a surface operator.
The computation for the first few terms
employs the Kac determinant at some lower levels.
We provide here the level-2 Kac determinant for reference.

The level-2 Kac determinant
is the determinant of the following Shapovalov matrix of level-2:
\begin{align}
Q_\Delta \,|_{|Y|=2}
=\left(\begin{array}{cc}4\Delta+\frac{c}{2} & 6\Delta 
\\6\Delta & 4\Delta(2\Delta+1)\end{array}\right).
\end{align}
Let us substitute the AGT parametrization $\Delta(\alpha)=(b+1/b)^2/4- \alpha^2$ and $c=1+6(b+1/b)^2$ 
into the Shapovalov matrix.
We then find that the Kac determinant can be factorized as follows:
\begin{align}
K_{{2}}(\Delta(\alpha))&=\det Q_\Delta \,|_{|Y|=2}\nonumber\\
&=-32\,\left( \alpha^2-\frac{1}{4}\left( b+\frac{1}{b}\right)^2\right)
\left( \alpha^2-\left( b+\frac{1}{2b}\right)^2\right)
\left( \alpha^2-\left( \frac{b}{2}+\frac{1}{b}\right)^2\right),
\end{align}
where the factors are related to the 2-instanton Nekrasov partition function
through the AGT relation. 
The determinant for $\Delta^{\prime}= \Delta \left(\alpha +b/4 \right) $ takes the following form
\begin{align}
K_{{2}} \left( \Delta \left(\alpha +b/4 \right)  \right) 
&=-\frac {1}{128}\,
\left( 4\,\alpha+3b+2\,{b}^{-1} \right) 
\left( 4\alpha -b-2\,{b}^{-1} \right)  
\left( 4\,\alpha -b-4\,{b}^{-1} \right)\nonumber\\
&\qquad\times
\left( 4\alpha -3b-2\,{b}^{-1} \right) 
\left( 4\,\alpha +5b+2\,{b}^{-1} \right) 
\left( 4\,\alpha +3b+4\,{b}^{-1} \right).
\end{align}
We need this factorized form in order to obtain the Nekrasov-like
expression for the irregular conformal block with the degenerate field.
The following formulae also play an important role in section 4.
\begin{align}
&\Delta\left(\alpha +\frac{b}{4} \right)=
-\left(\alpha -\frac{b}{4}-\frac{1}{2\,b} \right)
\left(\alpha +\frac{3b}{4}+\frac{1}{2\,b} \right),
\\
&\Delta\left(\alpha -\frac{b}{4} \right)=
-\left(\alpha -\frac{3\,b}{4}-\frac{1}{2\,b} \right)
\left(\alpha +\frac{b}{4}+\frac{1}{2\,b} \right),
\\
&\delta=\Delta\left(\alpha +\frac{b}{4} \right)-\Delta_{2,1}-\Delta\left(\alpha -\frac{b}{4} \right)
=-b\left(\alpha -\frac{3\,b}{4}-\frac{1}{2\,b} \right).
\end{align}

\subsection{Expansion coefficients}
The degenerate field on the descendant state has the following form of the expansion in the Verma module
\begin{align}
\label{Phi21descend}
\Phi_{2,1}(z)|\Delta, Y \rangle
=\sum_{Y^{\prime}}
z^{|Y^{\prime}|-|Y|+\delta}
\beta^{\,Y}_{\,Y^{\prime}}
\,|\Delta^\prime, Y^{\prime} \rangle.
\end{align}
It is easy to check this $z$-dependence of the expansion as follows:
let us expand $\Phi_{2,1}(z)|\Delta, Y \rangle$ in accordance with the level decomposition
\begin{align}
\Phi_{2,1}(z)|\Delta, Y \rangle
=\sum_{n}
|\Delta^\prime, Y, n ;z\rangle.
\end{align}
The commutation relation between a primary field and a Virasoro operator then implies
\begin{align}
L_0\Phi_{2,1}(z)|\Delta, Y \rangle
&=\sum_{n}
(\Delta^\prime+n)|\Delta^\prime, n ;z\rangle\nonumber\\
&=\left(z\frac{\partial}{\partial z}+\Delta_{2,1} \right)\Phi_{2,1}(z)|\Delta, Y \rangle
+\Phi_{2,1}(z)(\Delta+|Y|)|\Delta, Y \rangle.
\end{align}
This means that the $z$-dependence of the state is
\bea
z\frac{\partial}{\partial z}
|\Delta^\prime, Y, n ;z\rangle
 =(\delta-|Y|+n)
|\Delta^\prime, Y, n ;z\rangle ~
 \propto z^{\delta-|Y|+n},
\eea
where $\delta=\Delta^\prime-\Delta_{2,1}-\Delta$.

Take $Y =\bullet$ for example. Then the expansion (\ref{Phi21descend}) becomes
\begin{align}
\label{Phi21descend1}
\Phi_{2,1}(z)|\Delta \rangle
=\sum_{Y'}
z^{|Y'|+\delta}
\beta_{\,Y'}
\,|\Delta^\prime, Y' \rangle.
\end{align}
Here $\beta_{\,Y}$ denotes the coefficient for the empty Young diagram $\beta_{\,Y}=\beta^{\,\bullet}_{\,Y}$.
These expansion coefficients are determined by using the identity
\begin{align}
L_n\Phi_{2,1}(z)|\Delta \rangle
=
z^n\left(z\frac{\partial}{\partial z}+\Delta_{2,1}(n+1) \right)\Phi_{2,1}(z)|\Delta \rangle,
\end{align}
where $n>0$.
For $n=1$ this equation gives
\begin{align}
&2\Delta^\prime \beta_{\,1}
=
\delta+2\Delta_{2,1}\\
&(4\Delta^\prime+2)\beta_{\,1^2}
+3\beta_{\,2}
=(\delta+1+2\Delta_{2,1})\beta_{\,1},\\
&\qquad\qquad\qquad\qquad \cdots.\nonumber
\end{align}
The equation for $n=2$ implies
\begin{align}
&6\Delta^\prime\beta_{\,1^2}
+(4\Delta^\prime+c/2) \beta_{\,2}
=\delta+3\Delta_{2,1},\\
&\qquad\qquad\qquad\qquad \cdots.\nonumber
\end{align}
Thus, we get the first few coefficient for the degenerate field on the primary state $|\Delta\rangle$
\begin{align}
&\quad\,\,\beta_{\,1}=\frac{\Delta^\prime+\Delta_{2,1}-\Delta}{2\Delta^\prime},\\
&\left(\begin{array}{c}  \beta_{\,2}\\ \beta_{\,1^2}\end{array}\right)
=\,Q_{\Delta^\prime}^{-1}|_{|Y|=2}\cdot
\left(\begin{array}{c}  
\Delta^\prime+2\Delta_{2,1}-\Delta
\\ {(\Delta^\prime+\Delta_{2,1}-\Delta)(1+\Delta^\prime+\Delta_{2,1}-\Delta)} \end{array}\right).
\end{align}
Here $Q_\Delta$ is the Shapovalov matrix of level-2.
By using the relation, we obtain the following simple result which we use in section 4:
\begin{align}
\beta_{\,1^2}=
{\frac {1}{8\,b^2\, \left( a-b-\frac{1}{4b} \right)  \left( a-\frac{b}{2}-\frac{1}{4b} \right) }}.
\end{align}
The factors in the denominator relate to the instanton measure via the AGT relation.

For the first descendant state $Y =[1]$, the expansion (\ref{Phi21descend}) is
\begin{align}
\label{Phi21descend11}
\Phi_{2,1}(z)|\Delta, [1]\rangle
=\sum_{Y'}
z^{|Y'|-1+\delta}
\beta^{\,[1]}_{\,Y'}
\,|\Delta^\prime, Y' \rangle.
\end{align}
By using the commutation relation  and  (\ref{Phi21descend1}),
we have
\begin{align}
\Phi_{2,1}(z)|\Delta, [1]\rangle
&=L_{-1} \left( \,z^\delta\,|\Delta^\prime \rangle 
+z^{1+\delta}\beta_{\,1}\,|\Delta^\prime, [1] \rangle+\cdots\right) \nonumber\\
&\qquad\quad-\left( \delta\, z^{-1+\delta} \,|\Delta^\prime \rangle
+(1+\delta)z^\delta\beta_{\,1}\,|\Delta^\prime, [1] \rangle
+(2+\delta)z^{1+\delta}\beta_{\,1^2}\,|\Delta^\prime, [1^2] \rangle +\cdots\right)
\nonumber\\
&= -z^{-1+\delta}\delta \,|\Delta^\prime \rangle
+z^\delta(1-(1+\delta)\beta_{\,1})\,|\Delta^\prime, [1] \rangle\nonumber\\
&\qquad\quad+z^{1+\delta}(\beta_{\,1}-(2+\delta)\beta_{\,1^2})\,|\Delta^\prime, [1^2] \rangle
-z^{1+\delta}(2+\delta)\beta_{\,2}\,|\Delta^\prime, [2] \rangle
+\cdots .
\end{align}
In this way we can determine the expansion coefficients recursively:
\begin{align}
&\beta^{\,1}_{\,\bullet}=-\delta,\\
&\beta^{\,1}_{\,1}=1-(1+\delta)\beta_{\,1},\\
&\beta^{\,1}_{\,1^2}=\beta_{\,1}-(2+\delta)\beta_{\,1^2},\\
&\beta^{\,1}_{\,2}=-(2+\delta)\beta_{\,2}.
\end{align}

There are two Young diagrams with two boxes.
We study the diagram $Y =[1^2]$ first.
The expansion is
\begin{align}
\Phi_{2,1}(z)|\Delta, [1^2]\rangle
=\sum_{Y'}
z^{|Y'|-2+\delta}
\beta^{\,1^2}_{\,Y'}
\,|\Delta^\prime, Y' \rangle.
\end{align}
By using the commutation relation  and (\ref{Phi21descend11}),
we obtain the following expansion in the Verma module
\begin{align}
\Phi_{2,1}(z)|\Delta, [1^2]\rangle
&=\Phi_{2,1}(z)\cdot L_{-1}\,|\Delta, [1]\rangle\nonumber\\
&=L_{-1}\cdot \left(
 -z^{-1+\delta}\delta \,|\Delta^\prime \rangle
+z^\delta(1-(1+\delta)\beta_{\,1})\,|\Delta^\prime, [1] \rangle
+\cdots
\right)\nonumber\\
&\qquad\qquad-\frac{\partial}{\partial z}\, \left(
 -z^{-1+\delta}\delta \,|\Delta^\prime \rangle
+z^\delta(1-(1+\delta)\beta_{\,1})\,|\Delta^\prime, [1] \rangle
+\cdots
\right) \nonumber\\
&=z^{-1+\delta}\delta(-1+\delta) \,|\Delta^\prime \rangle
-z^{-1+\delta}\delta \,|\Delta^\prime , [1]\rangle
+
\cdots
.
\end{align}
Hence the expansion coefficients are given by
\begin{align}
&\beta^{\,1^2}_{\,\bullet}=\delta(\delta-1), \\
&\beta^{\,1^2}_{\,1}=-\delta,\\
&\beta^{\,1^2}_{\,1^2}=1-2(1+\delta)\beta_{\,1}+(2+\delta)\beta_{\,1^2},\\
&\beta^{\,1^2}_{\,2}=(1+\delta)(2+\delta)\beta_{\,2}.
\end{align}

We compute the expansion for $Y =[2]$ next:
\begin{align}
\Phi_{2,1}(z)|\Delta, [2]\rangle
=\sum_{Y'}
z^{|Y'|-2+\delta}
\beta^{\,2}_{\,Y'}
\,|\Delta^\prime, Y' \rangle.
\end{align}
By using the commutation relation  and  (\ref{Phi21descend1}) again,
we have
\begin{align}
\Phi_{2,1}(z)|\Delta, [2]\rangle
&=L_{-2}
\left( \,z^\delta\,|\Delta^\prime \rangle 
+z^{1+\delta}\beta_{\,1}\,|\Delta^\prime, [1] \rangle+\cdots\right)\nonumber\\
&\qquad\qquad -z^{-2}\left( z\frac{\partial}{\partial z} -\Delta_{2,1} \right)
\left( \,z^\delta\,|\Delta^\prime \rangle 
+z^{1+\delta}\beta_{\,1}\,|\Delta^\prime, [1] \rangle+\cdots\right)\nonumber\\
&=z^\delta|\Delta^\prime , [2]\rangle 
+z^{1+\delta}\beta_{\,1}\,|\Delta^\prime, [2\cdot1] \rangle+\cdots\nonumber\\
&\qquad\qquad -z^{-2}
\left( (\delta-\Delta_{2,1})z^{\delta}\ |\Delta^\prime \rangle 
+(1+\delta-\Delta_{2,1} )z^{1+\delta}\beta_{\,1}\,|\Delta^\prime, [1] \rangle+\cdots\right).
\end{align}
The expansion coefficients are given by
\begin{align}
&\beta^{\,2}_{\,\bullet}=\Delta_{2,1}-\delta,\\
&\beta^{\,2}_{\,1}=-\beta_{\,1}(1+\delta-\Delta_{2,1}),\\
&\beta^{\,2}_{\,1^2}=-\beta_{\,1^2}(2+\delta-\Delta_{2,1}),\\
&\beta^{\,2}_{\,2}=1-\beta_{\,2}(2+\delta-\Delta_{2,1}).
\end{align}

In section 4, we use these formulae for $\beta$'s to rewrite the irregular conformal block
as the ramified instanton partition function. 

\end{document}